\documentclass[reprint,prb]{revtex4-1}
\usepackage{amsmath,amssymb,amsmath,bm}
\usepackage{esint,color}
 \usepackage{graphicx,cancel}
\usepackage{mathtools}
\usepackage{empheq}
\usepackage{hyperref}
\usepackage{multirow}
\usepackage{nicefrac}
\usepackage{mathrsfs}
\usepackage[skins,theorems]{tcolorbox}
\tcbset{highlight math style={enhanced,
  colframe=red!60!black,colback=white,arc=4pt,boxrule=1pt}}
\makeatletter

\newsavebox{\@brx}
\newcommand{\llangle}[1][]{\savebox{\@brx}{\(\m@th{#1\langle}\)}%
  \mathopen{\copy\@brx\kern-0.5\wd\@brx\usebox{\@brx}}}
\newcommand{\rrangle}[1][]{\savebox{\@brx}{\(\m@th{#1\rangle}\)}%
  \mathclose{\copy\@brx\kern-0.5\wd\@brx\usebox{\@brx}}}
\makeatother

\newcommand{\tOmega}{ \tilde{\Omega} }
\newcommand{\dV}{d\tilde{V}}
\newcommand{\dS}{d\tilde{S}}

\newcommand{\x}[1]{{x^{#1}}}

\newcommand{\Jt}[2]{{\bf v}_#1^{\left( #2 \right)}}
\newcommand{\Jtoo}{{\bf v}}
\newcommand{\Jtt}[2]{{\bf v}_#1^{\left( #2 \right) \perp}}
\newcommand{\Jtl}[2]{{\bf v}_#1^{\left( #2 \right) \parallel}}
\newcommand{\Jtn}[2]{v_{#1 \, n}^{\left( #2 \right)} }

\newcommand{\Jl}[2]{{\bf u}_#1^{\left( #2 \right)}}
\newcommand{\Jloo}{{\bf u}}
\newcommand{\Jlt}[2]{{\bf u}_#1^{\left( #2 \right) \perp}}
\newcommand{\Jll}[2]{{\bf u}_#1^{\left( #2 \right) \parallel}}
\newcommand{\Jln}[2]{ u_{n,#1}^{\left( #2 \right)}}

\newcommand{\Jlo}{{\bf j}^{\parallel}}
\newcommand{\Jlnn}[1]{j_{n,#1}^{\parallel}}
\newcommand{\gammalo}[1]{\chi^{\parallel}_{#1}}

\newcommand{\Jto}{{\bf j}^{\perp}}
\newcommand{\gammato}[1]{\kappa^{\perp}_{#1}}

\newcommand{\gammat}[2]{\kappa^{\left( #2 \right)}_{#1}}
\newcommand{\gammal}[2]{\chi^{\left( #2 \right)}_{#1}}

\newcommand{\ahk}[3]{\alpha^{\left( #3 \right)}_{#1,#2} }
\newcommand{\bhk}[3]{\beta^{\left( #3 \right)}_{#1,#2} }

\newcommand{\nim}{n_i}

\newcommand{\re}[1]{\text{Re}\left\{ #1 \right\}}
\newcommand{\im}[1]{\text{Im}\left\{ #1 \right\}}

\newcommand{\rt}{\tilde{\mathbf{r}}}
\newcommand{\rbt}{\tilde{\mathbf{r}}}
\newcommand{\rp}{\left( \tilde{\mathbf{r}} \right)}
\newcommand{\rpp}{\left( \tilde{\mathbf{r}}' \right)}

\newcommand{\Jls}{\left\{ {\bf u}_h \left( \rt \right)\right\}}
\newcommand{\Jts}{\left\{ {\bf v}_h \left( \rt \right)\right\}}

\newcommand{\go}{g_0 \left( \tilde{\mathbf{r}} - \tilde{\mathbf{r}}'\right)}
\newcommand{\n}{\hat{\mathbf{n}}}
\newcommand{\Ei}{\mathbf{E}_{inc} \left( \mathbf{r} \right)}
\newcommand{\W}{ \mathbf{W} \left( \tilde{\mathbf{r}}' \right)}

\newcommand{\tnabla}{\tilde{\nabla}}
\newcommand{\Who}{ \mathbf{W}_k^\perp }

\newcommand{\Qtd}[1]{\text{Q}_{#1}^{\perp \text{nr}} }
\newcommand{\Qtr}[1]{\text{Q}_{#1}^{\perp \text{r}} }
\newcommand{\Qtt}[1]{\text{Q}_{#1}^{\perp}}

\newcommand{\Qlr}[1]{\text{Q}_{#1}^{\parallel \text{r}} }
\newcommand{\Qld}[1]{\text{Q}_{#1}^{\parallel \text{nr}} }
\newcommand{\Qlt}[1]{\text{Q}_{#1}^{\parallel}}

\newcommand{\Qh}[1]{\hat{\text{Q}}_{#1}}

\newcommand{\deltar}{\Delta\rt}

\newcommand{\jtTEd}[1]{\mathbf{j}^{\text{TE}\perp}_{o1#1}}
\newcommand{\jtTMd}[1]{\mathbf{j}^{\text{TM}\perp}_{e1#1}}

\newcommand{\jtTE}[1]{\mathbf{j}^{\text{TE}\perp}_{\substack{e \\ o}#1}}
\newcommand{\jtTM}[1]{\mathbf{j}^{\text{TM}\perp}_{\substack{e \\ o}#1}}

\newcommand{\kappaTE}[2]{\kappa_{#2}^{\text{TE} \, #1}}
\newcommand{\kappaTM}[2]{\kappa_{#2}^{\text{TM}\,  #1}}

\newcommand{\PM}[1]{\mathbf{P}_{\text{M}|#1}^\perp}

\newcommand{\QE}[1]{\tensor{\mathbf{Q}}_{\text{E}|#1}^\parallel}
\newcommand{\QM}[1]{\tensor{\mathbf{Q}}_{\text{M}|#1|ij}^\perp}
\newcommand{\TE}[1]{\mathbf{P}_{\text{E2}|#1}^\perp}
\newcommand{\PE}[1]{\mathbf{P}_{\text{E}|#1}^\parallel}

\begin{document}

\title{Resonance frequency and radiative Q-factor \\ of 
plasmonic and dielectric modes  \\ of small objects}

\author{Carlo Forestiere}
\affiliation{ Department of Electrical Engineering and Information Technology, Universit\`{a} degli Studi di Napoli Federico II, via Claudio 21,
 Napoli, 80125, Italy}
\author{Giovanni Miano}
\affiliation{ Department of Electrical Engineering and Information Technology, Universit\`{a} degli Studi di Napoli Federico II, via Claudio 21,
 Napoli, 80125, Italy}
 \author{Guglielmo Rubinacci}
\affiliation{ Department of Electrical Engineering and Information Technology, Universit\`{a} degli Studi di Napoli Federico II, via Claudio 21,
 Napoli, 80125, Italy}

\begin{abstract}
The electromagnetic scattering resonances of a non-magnetic object much smaller than the incident wavelength in vacuum can be either described by the electroquasistatic approximation of the Maxwell's equations if its permittivity is negative, or by the magnetoquasistatic approximation if its permittivity is positive and sufficiently high. Nevertheless, these two approximations fail to correctly account for the frequency shift and the radiative broadening of the resonances when the size of the object becomes comparable to the wavelength of operation. In this manuscript, the radiation corrections to the electroquasistatic and magnetoquasistatic resonances of arbitrarily-shaped objects are derived, which only depend on the quasistatic current modes.  Then, closed form expressions of the frequency-shift and the radiative Q-factor of both plasmonic and dielectric modes of small objects are introduced, where the dependencies on the material and the size of the object are factorized. In particular, it is shown that the radiative Q-factor explicitly depends on the multipolar components of the quasistatic modes.
\end{abstract}

\maketitle

\section{Introduction}

There exist two mechanisms through which a non-magnetic homogeneous object, assumed small compared to the incident wavelength in vacuum, may resonate.

The first resonance mechanism occurs in small metal nanoparticles with negative permittivity, and it arises from the interplay between the energy stored in the electric field and the kinetic energy of the free electrons of the metal.  When the object is very small compared to the wavelength in vacuum, these resonances are well described by the electroquasistatic approximation of the Maxwell’s equations \cite{Fredkin:03,Bergman:03,Li:03,Wang:06,klimov2014nanoplasmonics} and associated to the negative values of permittivity in correspondence of which source-free solutions exist. However, it is known that, as the size of the object becomes comparable to the incident wavelength, this approximation is unable to describe the radiative shift and broadening of these resonances.  

The second resonance mechanism occurs in small objects of high and positive permittivity, and it arises from the interplay between the polarization energy stored in the dielectric and the energy stored in the magnetic field. Manifestation of this kind of resonance can be found at microwaves  \cite{richtmyer1939dielectric,kajfez1998dielectric,Long:83,Mongia:94}, and at optical \cite{garcia2011strong,evlyukhin2012demonstration,kuznetsov2012magnetic,Kuznetsov:16,PhysRevLett.119.243901} frequencies.  When the object is very small compared to the free-space wavelength, and the permittivity very high, these resonances are well-described by the magnetoquasistatic approximation of the Maxwell’s equations \cite{Forestiere:20}, where the normal component of the displacement current density field vanishes on the surface of the particle \cite{VanBladel:75a,Forestiere:20}. In particular, these resonances are associated with the eigenvalues of the magnetostatic integral operator expressing the vector potential in terms of the displacement current density  \cite{Forestiere:20}. Unfortunately, when the permittivity of the dielectric material is only moderately high, as happens for instance in the visible spectral range \cite{garcia2011strong,evlyukhin2012demonstration,kuznetsov2012magnetic,Kuznetsov:16,PhysRevLett.119.243901}, the size of the object has to be comparable to the incident wavelength to trigger a resonant response. In this scenario the magnetoquasistatic approximation is unable to describe the frequency shift and the broadening of these resonances.

In light of these observations, to describe the electromagnetic resonances of objects of size comparable to the incident wavelength, one may be tempted to abandon the {\it quasistatic} approximation altogether and then turn to the full-Maxwell's equations, using one of several possible spectral approaches, including the quasi-normal-modes  \cite{muljarov2011brillouin,Lalanne:18}, the characteristic modes \cite{Garbacz:71,1140154}, or the material-independent-modes \cite{Bergman80,Forestiere:16}. There are certainly some advantages in doing so, including the fact that the full-wave formulations would enable, given unlimited computational resources, the treatment of objects of any size. In fact, scattering resonances have been already investigated  by considering full-wave eigenvalue problems based on volume  (e.g.   \cite{Zheng:13,deLasson:13,ForestiereAP}), surface  (e.g. \cite{Makitalo:14,Powell:14,Bernasconi:16,Powell:17,Forestiere2D}), and line integral formulations of the Maxwell's equations \cite{Forestiere1D}, differential formulations (e.g.  \cite{Bai:13,PhysRevA.101.011803}), Mie Theory (e.g. \cite{muljarov2011brillouin,Forestiere:16,Forestiere_2017,Pascale:19}), for a recent review see Ref. \cite{Lalanne:18}. However, the resulting resonance frequencies, Q-factors, and resonant modes depend on the morphology,  material,  and size of the object. A change of any of these parameters would require an entirely new calculation. These dependencies are buried below the computational layer, and cannot be factorized.  

Closed form expressions of the Q-factor and the frequency shift of both plasmonic and dielectric modes, where the dependencies on the material and size of the object are factorized, are highly desirable. They would enable the classification of the resonances, and  facilitate their engineering \cite{doi:10.1063/1.5094188}, including the coupling with emitters \cite{Zambrana:15,Krasnok:16,regmi2016all,Sapienza:19}, because they could be used as a target for the design \cite{Wiecha:17,bonod2019evolutionary}.  Moreover, in many applications, the size of metal or dielectric objects does not exceed the free-space wavelength of operation \cite{Kuznetsov:16}. These are powerful incentives to pursue the extension of the two quasi-static scattering limits to include radiation effects.

In the literature there already exist closed form expressions for the resonance frequency shift and Q-factors in few scenarios.  

For electrically small antennas, physical limitations on the Q-factor have been the subject of numerous papers\cite{harrington1960effect,collin1964evaluation,hansen1981fundamental,mclean1996re,yaghjian2005impedance,Gustafsson:07}, starting from the classical works of H. A. Wheeler \cite{wheeler1947fundamental} and L. J. Chu \cite{chu1948physical}.

For plasmonic resonators, Mayergoyz et. al derived the second order correction to electroquasistatic eigenvalues \cite{Mayergoyz:05} , starting from Maxwell's equations in differential form.  Wang and Shen derived the expression of the Q-factor of a plasmonic mode, when the non-radiative losses are dominant \cite{Wang:06}. To the authors' best knowledge the derivation of a general expression for the Q-factor of plasmonic modes when radiative losses are dominant \cite{PhysRevLett.97.263902} is still missing.

For  high-index dielectric resonators, Van Bladel introduced closed form expressions for the Q-factor of the magnetic dipole mode, and provided the Q-factor of specific higher order modes of a rotationally symmetric  object \cite{VanBladel:75a}, considering an asymptotic expansion of the Maxwell's equation in differential form in terms of the inverse of the index of refraction.  Following Van Bladel's work,  De Smedt derived the frequency shift and Q-factor of a {\it rotationally} symmetric ring resonator \cite{DeSmedt84}. General expressions for both the frequency shift and radiative factor haven't been derived yet.

In this paper, the radiation corrections for both the electroquasistatic and magnetoquasistatic resonances and resonant modes of arbitrarily-shaped non-magnetic homogeneous and isotropic objects are introduced, using an integral formulation of the Maxwell's equations and treating as a small parameter.

 
  It is demonstrated that, in the scattering from small objects, the relative resonance frequency shift of any mode (with respect to the quasistatic resonance position) is a quadratic function of the size parameter at the quasistatic resonance, whose prefactor depends on the ratio between the second order correction and the quasistatic eigenvalue. Furthermore, the {\it radiative} Q-factor is an inverse power function of the size parameter whose exponent is the order $n_i$ of the first non-vanishing imaginary correction, while the prefactor is the ratio between the quasistatic eigenvalue and the  $n_i$-th order imaginary correction, which explicitly depends on the multipolar components of the quasistatic mode.

This manuscript is organized as follows. First, the scattering resonances in the two quasistatic regimes  are briefly summarized in Sec. \ref{sec:QS}. Then, in Sec. \ref{sec:EM}, the full-wave scattering problem is formulated and an eigenvalue problem governing the scattering resonances is introduced. This eigenvalue problem is solved perturbatively in Secs. \ref{sec:Plasmonic} and \ref{sec:dielectric}, starting from the electroquasistatic and magnetoquasistatic limits, treating the size parameter as a small parameter. Collecting same-order terms, closed form radiation corrections are found. In Sec. \ref{sec:Q},  the frequency shift and the Q-factor are obtained as a function of these radiation corrections.  In Sec. \ref{sec:Tables}, the catalogues of plasmonic and photonic resonances are introduced. They constitute a synthetic classification of the modes of a homogeneous non-magnetic objects, and  depend  only on its morphology, but not on its size, material, and frequency of operations. This classification may help the description of the elementary building blocks of the nano-circuitry envisioned by Engheta et al. in Ref. \cite{Engheta1698}.
Eventually, the introduced formalism is validated by investigating the resonance frequency and Q-factors in the scattering response of metal and dielectric objects of size comparable to the incident wavelength. 

\section{Resonances in the Quasistatic Regime} 
\label{sec:QS}
A homogeneous, isotropic, non-magnetic, linear material occupies a volume  $\Omega$, of characteristic linear dimension $l_c$, bounded by a closed surface $\partial \Omega$  with an outward-pointing normal $\n$. The material has relative permittivity $\varepsilon_R \left( \omega \right)$, and it is surrounded by vacuum.  There exist two mechanisms through this object may resonate in the quasistatic regime. \cite{Forestiere:20}

\subsection{Electroquasistatic resonances}
\label{sec:EQS}
The first resonance mechanism is the electroquasistatic resonance, occurring in metals (more in general, in objects whose dielectric permittivity has a negative real part) where the induced electric charge plays a central role. These resonances are associated with the eigenvalues $\gammalo{h}$ of the electrostatic integral operator $\mathscr{L}_e$ that gives the electrostatic field as a function of the surface charge density \cite{mayergoyz2013plasmon}:
\begin{equation}
\gammalo{h} \, \mathscr{L}_e \left\{ \Jlo_h \right\} =  \Jlo_h,
\label{eq:EQSproblem}
\end{equation}
the expression of $\mathscr{L}_e$ is
\begin{equation}
  \mathscr{L}_e \left\{ \mathbf{W} \right\} = - \tilde{\nabla} \oiint_{ \partial \tilde{\Omega}} \go \mathbf{W} \rpp \cdot \n \rpp d\tilde{S}',
 \label{eq:EQSop}
\end{equation}
where $g_0 \left( \rt - \rt' \right) = \frac{1}{4\pi \left| \rt - \rt' \right| }$ is the static Green function in vacuum. In Eq. \ref{eq:EQSop} the spatial coordinates have been normalized by $l_c$, i.e. $\rt = \mathbf{r}/l_c$, $\tOmega$ is the corresponding scaled domain, $\partial \tOmega$ is the boundary of $\tOmega$,  and $\tnabla$ is the scaled gradient operator. 

The quasistatic oscillations, represented by the electroquasistatic  (EQS) current modes $\Jlo_h$, arise from the interplay between the  energy stored in the electric field and the kinetic energy of the free electrons in the metal. The spectrum of the operator $\mathscr{L}_e$ is discrete \cite{Mayergoyz:05,mayergoyz2013plasmon}. Each EQS mode $\Jlo_h$ is characterized by a real and negative eigenvalue $\gammalo{h}$, which is size-independent \cite{Mayergoyz:05}. The modes $\left\{ \Jlo_h \right\}_{h \in \mathbb{N}}$ are {\it longitudinal} vector fields: they are both curl-free and div-free within the object, but have non-vanishing normal component to the object surface \cite{Mayergoyz:05,mayergoyz2013plasmon}. This normal component is related to the induced surface charge density on $\partial \tOmega$, and  satisfies the {\it charge-neutrality} condition, i.e.:
\begin{equation}
\oiint_{ \partial \tilde{\Omega}}   \Jlo_h \left( \rbt' \right) \cdot \tilde{\mathbf{n}} \rpp \, d\tilde{S}' = 0.
\label{eq:ChargeNeutrality}
\end{equation}
Moreover, the EQS modes are orthonormal:
\begin{equation}
  \langle \Jlo_h, \Jlo_k \rangle_{\tOmega} = \delta_{h,k}
  \label{eq:EQSorthonorm}
\end{equation}
accordingly to the scalar product
\begin{equation}
   \langle  \mathbf{f} , \mathbf{g}  \rangle_{\tOmega} = \iiint_{\tOmega}  \mathbf{f}^*  \left( \rbt \right) \cdot  \mathbf{g} \left( \rbt \right) d \tilde{V}.
 \label{eq:ScalarProd}
\end{equation}
Under the normalization \ref{eq:EQSorthonorm}, the electrostatic energy of the $h$-th EQS current mode is 
\begin{equation}
 \mathscr{W}_e \left\{ \Jlo_h \right\} =  \frac{1}{2 \varepsilon_0} \frac{1}{\left( -\chi^\parallel_h \right)}.
\end{equation}

The electric dipole moment $\mathbf{P}_{\text{E},h}^\parallel$  of the EQS mode $\Jlo_h$ is defined accordingly to Eq. \ref{eq:Dipole} of the Appendix \ref{sec:Multipoles}. If the mode  $\Jlo_h$ exhibits a vanishing electric dipole moment, i.e.: 
\begin{equation}
  \left\| \mathbf{P}_{\text{E},h}^\parallel \right\| = 0,
\end{equation}
it is called {\it dark}, {\it bright} otherwise.

\subsection{Magnetoquasistatic resonances}
\label{sec:MQS}
The second resonance mechanism is the magnetoquasistatic resonance, occurring in dielectric objects with high and positive permittivity, where  the  displacement current density field plays a central role.  These resonances are associated with the eigenvalues $\gammato{h} $ of the magnetostatic integral operator $\mathscr{L}_m$ that gives the vector potential as a function of the current density \cite{Forestiere:20}:
\begin{equation}
\gammato{h} \, \mathscr{L}_m \left\{ \Jto_h \right\} =  \Jto_h,
\label{eq:MQSproblem}
\end{equation}
with
\begin{equation}
\left. \Jto_h \rp \cdot \n \rp  \right|_{\partial \tOmega} = \mathbf{0} \qquad \forall \rt \in \partial \tOmega,
\end{equation}
the expression of $\mathscr{L}_m$ is 
\begin{equation}
\mathscr{L}_m \left\{  \Who \right\} \left( \rt \right)  = \iiint_{\tilde{\Omega}} \go \Who \rp  d\tilde{V}'.
\label{eq:MQSoperator}
\end{equation}
Equation \ref{eq:MQSproblem} holds in the weak form in the functional space equipped with the inner product \ref{eq:ScalarProd}, and constituted by the {\it transverse} vector fields which are div-free within $\tOmega$ and having zero normal component to $\partial \tOmega$.

The  quasistatic oscillations represented by the magnetoquasistatic (MQS) current density modes $\Jto_h$  arise from the interplay between the polarization energy stored in the dielectric and the energy stored in the magnetic field \cite{Forestiere:20}. The spectrum of the magnetoquasistatic operator \ref{eq:MQSproblem} is discrete, too \cite{Forestiere:20}. The MQS current mode $\Jto_h$  is characterized by a real and positive eigenvalue $\gammato{h}$, which is size-independent. The current modes $\left\{ \Jto_h \right\}_{h \in \mathbb{N}}$ are transverse modes: they have a non-zero curl within the object, but are divergence-free and have a vanishing normal component on the object surface \cite{Forestiere:20}.  Each current mode $\Jto_h$ has zero electric dipole moment, namely:
\begin{equation}
   \iiint_{\tOmega} \Jto_h \rpp d\tilde{V} = 0.
   \label{eq:ZeroAverage}
\end{equation}
Furthermore, the MQS current density modes are orthonormal:
\begin{equation}
   \langle \Jto_h | \Jto_k \rangle = \delta_{h,k}.
    \label{eq:MQSorthonorm}
\end{equation}
Under this normalization, the magnetostatic energy of the $h$-th electroquasistatic current mode is 
\begin{equation}
 \mathscr{W}_m \left\{ \Jto_h \right\} =  \frac{\mu_0}{2} \frac{1}{\kappa_h^\perp}.
\end{equation}

The current mode $ \Jto_h$ generates a vector potential
\begin{equation}
  \mathbf{A} \left\{ \Jto_h \right\} \rp = \frac{\mu_0}{ 4 \pi}  \iiint_{\tOmega}   \go \Jto_h \rpp \dV'.
\end{equation}
Among the MQS current modes, there exists a subset of modes generating a vector potential $\mathbf{A} \left\{ \Jto_h \right\}$ with zero normal component to $\partial \tOmega$, i.e.
\begin{equation}
    \n \rp \cdot \mathbf{A} \left\{ \Jto_h \right\} \rp  = 0   \qquad \forall \rt \in \partial \tOmega;
    \label{eq:Amode}
\end{equation}
namely $\mathbf{A} \left\{ \Jto_h \right\} \rp$ is a transverse field. In this manuscript, a MQS mode belonging to this subset is called $\mathbb{A}^\perp$-mode. The $\mathbb{A}^\perp$-modes are also solution of the problem \ref{eq:MQSproblem}  in a strong form (in the space of square integrable vector fields).

The longitudinal set of EQS current modes $\left\{ \Jlo_h \right\}_{h \in \mathbb{N}}$ and the transverse set of MQS $\left\{ \Jto_h \right\}_{h \in \mathbb{N}}$ modes are orthogonal accordingly to the scalar product \ref{eq:ScalarProd}, and together are a complete basis of the vector space of square integrable divergence-free vector fields in $\Omega$.

\section{Electromagnetic modes}
\label{sec:EM}
The full-wave scattering problem can be formulated by considering as unknown the current density field $\mathbf{J}$ induced in the object. This current density  particularizes into conduction current in metals at frequencies below interband transitions, polarization current in dielectrics, sum of conduction and polarization currents in metals in the frequency ranges where interband transitions occur. The object is illuminated by a time harmonic electromagnetic field $\re{\Ei e^{i \omega t}}$. In the frequency domain, the field ${\bf J} \left( {\bf r} \right)$ is related to the electric field $\bf E \left( {\bf r} \right)$ by ${\bf J} \left( {\bf r} \right) = i \omega \varepsilon_0 \chi \left( \omega \right) {\bf E}  \left( {\bf r} \right)$  where  $\chi \left( \omega \right) = \left( \varepsilon_R \left( \omega \right) - 1 \right)$ is the electric susceptibility of the object,  and  $\varepsilon_0$  is the vacuum permittivity. Both the vector field $\bf E$  and ${\bf J}$ are divergence-free in the region $\Omega$ occupied by the object due to the homogeneity and isotropy of the material. The induced current density is solution of the full-wave volume integral equation \cite{jin2011theory,van2007electromagnetic,hanson2013operator}:
\begin{multline}
\frac{ {\bf J} \left( {\bf r} \right)}{i \omega \varepsilon_0 \chi} = - \frac{1}{i \omega \varepsilon_0}
   \nabla \oiint_{\Omega} g \left( {\bf r} - {\bf r}'  \right) {\bf J} \left( {\bf r}' \right) \cdot \n \left( {\bf r}' \right) d S' \\ - i \omega \mu_0 \iiint_{\Omega} g \left( {\bf r} - {\bf r}'  \right)  {\bf J} \left( {\bf r}' \right)  dV' + \Ei \qquad \forall  {\bf r} \in \Omega,
 \label{eq:VIE}
\end{multline}
where $\mu_0$  is the vacuum permeability,  $g \left( \mathbf{r} \right) = e^{-i k_0 r} / 4 \pi r $  is the Green function in vacuum, $k_0 = \omega/ c_0$   and $c_0 = 1 / \sqrt{\varepsilon_0 \mu_0}$.
The surface and volume integrals represent the contributions to the induced electric field of the scalar and vector potentials, respectively. 
Then, equation \ref{eq:VIE} is rewritten as  \cite{ForestiereAP}
\begin{equation}
  \frac{ \mathbf{J} \left( \rt \right) }{\chi} - \mathscr{L} \left\{ \mathbf{J} \right\} \left( \rt \right) = i \omega \varepsilon_0 \mathbf{E}_{inc} \left( \rt \right) \quad \forall \rt\in \tilde{\Omega},
 \label{eq:VIEnorm}
\end{equation}
where the spatial coordinates are normalized as $\rt = \mathbf{r}/l_c$,
\begin{multline}
 \mathscr{L} \left\{ \mathbf{W} \right\} \left( \rt \right) =
  - \tnabla \oiint_{\partial \tilde{\Omega}} g \left( \rbt - \rbt', x \right)  \W \cdot \n \rpp d\tilde{S}' \\\ + x^2 \iiint_{\tilde{\Omega}} g \left( \rbt - \rbt', x \right) \W d\tilde{V}',
   \label{eq:Loperator}
\end{multline}
$\tilde{\Omega}$ is the scaled domain,  $\partial{\tOmega}$ is boundary of $\tOmega$, 
$\tilde \nabla$  is the scaled gradient operator, $x$ is the size parameter $x = 2 \pi l_c / \lambda$, and $g \left( \rt - \rt', x \right)$ is the Green function in vacuum
\begin{equation}
      g \left( \rt - \rt', x \right) = \frac{1}{4 \pi} \frac{e^{-i x \left| \rbt - \rbt' \right|}}{\left| \rbt - \rbt' \right|} =  \frac{1}{4 \pi} \frac{e^{-i x \deltar}}{\deltar},
\end{equation}
and $\deltar = \left| \rbt - \rbt' \right|$.  

The spectral properties of the linear operator $\mathscr{L}$ play a very important role in any resonant scattering mechanism. Since $\mathscr{L}$ is compact its spectrum is countable infinite. The operator $\mathscr{L}$ is symmetric but not self-adjoint. For any value of the size parameter $x$  its eigenvalues are complex with positive imaginary part. The eigenfunctions corresponding to two different eigenvalues are not orthogonal in the usual sense: they are bi-orthogonal  \cite{Bergman80,Forestiere:16}.

The eigenvalue problem  \cite{ForestiereAP}
\begin{equation}
\mathscr{L} \left\{ {\bf j}_k \right\} = \frac{1}{\gamma_h} \, {\bf j}_k
\label{eq:EigVIE}
\end{equation}
splits into the two eigenvalue problems \ref{eq:EQSproblem} and \ref{eq:MQSproblem} (see Sections \ref{sec:EQS} and Section \ref{sec:MQS}) in the quasi-static regime $x \ll 1$ (small object).  This fact was already shown for 2D objects in Ref. \cite{Forestiere2D} and for 3D objects in Ref. \cite{Forestiere:20}.
The eigenfunctions of $\mathscr{L}$ that in the limit $x \rightarrow 0$  tend to the EQS modes  are indicated with $\Jls$ and the corresponding eigenvalues are indicated as  $\left\{\chi_h\right\}$. These eigenfunctions are called {\it  plasmonic modes}. Dually, the set of eigenfunctions of $\mathscr{L}$ that in the limit $x \rightarrow 0$  tend to the MQS modes are indicated with $\Jts$  and the corresponding eigenvalues are indicated with $\kappa_h/x^2 $.  Although in the limit $x \rightarrow 0$, the eigenvalues $\kappa_h/x^2$ diverge, the quantities $\kappa_h$ remain constant. These eigenfunctions are called {\it dielectric modes}. 
Forestiere and Miano et al. in Ref. \cite{Forestiere:16,Forestiere_2017} used the adjectives {\it plasmonic} and  {\it photonic} mode instead of  {\it plasmonic} and {\it dielectric} mode to identify the same two sets, while in Ref. \cite{Pascale:19} the authors called them {\it longitudinal} and {\it transverse} modes. All these nomenclatures are equivalent.  It was shown that this two sets of modes, even if this distinction is made in the long-wavelength regime, remain well distinguishable and have different properties even in the full-wave regime \cite{Forestiere:16,Forestiere_2017,Forestiere2D}.

The union of the two sets  $\Jls$ and $\Jts$ is a basis for the unknown current density field in equation \ref{eq:VIE}. Its solution is expressed as
\begin{multline}
   {\bf J} \rp = \chi \left[ \sum_{h=1}^\infty \frac{ \chi_h }{ \chi_h - \chi } \langle   \Jloo_h^*, {\bf E}_{inc} \rangle_{\tOmega} \; \Jloo_h \rp
    \right. \\ \left. + \sum_{h=1}^\infty \frac{ \kappa_h }{ \kappa_h - \chi x^2 } \langle  \Jtoo_h^*, {\bf E}_{inc} \rangle_{\tOmega} \; \Jtoo_h \rp \right]
\label{eq:MIMexpansion}
\end{multline}
where both the set of modes $\Jls$  and $\Jts$  are normalized, $\langle  \Jloo_h ^*, \Jloo_h \rangle = 1$ and $\langle  \Jtoo_h^*,\Jtoo_h \rangle = 1$  for any  $h$. This expansion is very useful because it separates the dependence on the material from the dependence on the geometry \cite{Bergman80,Forestiere:16,Forestiere2D,ForestiereAP,Forestiere1D}, and has been used in different contexts \cite{Forestiere_2017,forestiere2019directional,PhysRevB.101.155401}. 

In the next two sections, we develop a perturbation theory to evaluate the plasmonic and dielectric resonances and resonant modes of an object with arbitrary shape and size parameter $x \lessapprox 1 $,  by starting from the corresponding modes in the quasistatic regime.


\section{Plasmonic Resonances}
\label{sec:Plasmonic}
To evaluate the plasmonic resonances of small particles, it is convenient to recast the eigenvalue problem \ref{eq:EigVIE} as
\begin{multline}
 -  \Jloo \rp   - {\chi} \, \tnabla \oiint_{\partial \tilde{\Omega}} g \left( \rbt - \rbt', x \right) \Jloo \rpp  \cdot \n \rpp \,  d\tilde{S}' + \\ + {\chi} \, x^2  \, \iiint_{\tilde{\Omega}} g \left( \rbt - \rbt' , x \right) \Jloo \rpp \dV' = \mathbf{0} \quad \forall \rt\in \tilde{\Omega} .
 \label{eq:VIE0}
\end{multline}
When the free-space wavelength $\lambda = 2 \pi c_0 / \omega$ is large in comparison with the characteristic dimension $l_c$, the size parameter $x$ can be treated as a small parameter, and the Green function $g \left( \rbt - \rbt' , x \right)$,  the current mode $\Jloo_h$, and the eigenvalue $\chi_h$  can all be expanded in terms of $x$ in the neighborhood of the EQS resonance with eigenvalue $\gammalo{h}$ and mode $\Jlo_{h}$:
\begin{align}
\label{eq:ChiExp}
 \chi_h & = \gammalo{h} + \gammal{h}{1}x +  \gammal{h}{2}x^2+\gammal{h}{3}x^3 + \ldots = \sum_{k=0}^{\infty}  \gammal{h}{k} x^k \, , \\
\label{eq:JExp}
 \Jloo_h & =   \Jlo_{h} + \Jl{h}{1} x + \Jl{h}{2} \x{2}  + \Jl{h}{3} \x{3} + \dots  =    \sum_{k=0}^{\infty} \Jl{h}{k} \x{k} \, ,
\end{align}
\begin{multline}
 g \left( \rt - \rt', x \right) = \frac{1}{4\pi} \left( \deltar^{-1}  - i x -\frac{\deltar}{ 2 !}x^2 + i \frac{\deltar^2}{3!} x^3 + \ldots \right) \\ =  \frac{1}{4\pi}  \sum_{k=0}^\infty (-i)^k \frac{\deltar^{k-1}}{k!} \x{k} \, .
 \label{eq:GExp} 
\end{multline} 
By using Eqs. \ref{eq:ChiExp}, \ref{eq:JExp}, and \ref{eq:GExp}, Eq. \ref{eq:VIE0} becomes
\begin{widetext}
\begin{multline}
  - {4 \pi} \sum_{k=0}^\infty \Jl{h}{k} \x{k}  -  \left( \sum_{k=0}^\infty \gammal{h}{k} \x{k} \right)  \tnabla  \oiint_{\partial \tOmega}  \left( \sum_{k=0}^\infty (-i)^k \frac{\deltar^{k-1}}{k!} \x{k}
    \right) \left( \sum_{k=0}^\infty \Jln{h}{k} \x{k} \right) d\tilde{S}' +  \\ \left( \sum_{k=0}^\infty \gammal{h}{k} \x{k} \right) \iiint_{\tOmega}  \left( \sum_{k=0}^\infty (-i)^k \frac{\deltar^{k-1}}{k!} \x{k}
    \right) \left( \sum_{k=0}^\infty \Jl{h}{k} \x{k+2} \right) \dV'  = 0, \qquad \forall \mathbf{r} \in \tOmega,
   \label{eq:VIEeqs} 
\end{multline}
\end{widetext}
where $\left. \Jln{h}{k} = \Jl{h}{k} \cdot \mathbf{n}\right|_{\partial \tOmega}$ and $\gammal{h}{0} = \gammalo{h}$ and $\Jl{h}{0}= \Jlo_h$.
In the Supplemental Material  all the details of the derivation of radiation corrections for plasmonic resonances are reported \cite{SI}. In the following, the EQS current modes are normalized accordingly to Eq. \ref{eq:EQSorthonorm}, i.e. $\left\| \Jlo_h \right\|=1$, $\forall h$.

Matching the first-order terms in Eq. \ref{eq:VIEeqs}, it is obtained that the first order corrections vanish regardless of the object's shape:
\begin{align}
\gammal{h}{1} & =0, \\
\label{eq:Chi1}
 \Jl{h}{1} \rp & ={\bf 0} \qquad \forall {\bf r} \in \tOmega.
\end{align}

Collecting the second order terms in Eq. \ref{eq:VIEeqs}, and applying the normal solvability condition of Fredholm integral equations \cite{mikhlin1970mathematical,Kantorovich:82}, the second order correction $\gammal{h}{2}$ is derived 
\begin{widetext}
 \begin{equation}
 \gammal{h}{2}  =
     -   \left(\gammalo{h}\right)^2  \frac{1}{4 \pi }  \left(  
      \oiint_{\partial \tOmega}   \Jlnn{h} \rp  \oiint_{\partial \tOmega}   \frac{\left| \rbt - \rbt' \right|}{ 2}\Jlnn{h} \left( \rbt' \right)
 \dS' \dS  +   \iiint_{\tOmega}  \Jlo_{h} \rp \cdot \iiint_{\tOmega} \frac{ \Jlo_{h} \rpp}{{\left| \rbt - \rbt' \right|}} \, \dV' \dV \right),
   \label{eq:Chi2b} 
\end{equation}
\end{widetext}
where the scalar field $\Jlnn{h} \left( \rbt \right) = \left. \Jlo_{h} \rp  \cdot \hat{\mathbf{n}} \rp \right|_{\partial \tOmega}$ is defined on the object's surface $\partial \tOmega$. Accordingly to Eq. \ref{eq:Chi2b}, $\gammal{h}{2}$ is real. Moreover, the first term in parenthesis in Eq. \ref{eq:Chi2b} (starting from the left) originates from the radiative self-interaction of the surface charge density associated to the EQS current mode through the scalar potential. The second term is instead proportional to the magnetostatic energy of the current mode $\Jlo_{h}$.
A second order correction to the EQS modes has been already derived in Ref. \cite{Mayergoyz:05} by expanding the Maxwell's equation in differential form. It will be demonstrated in Eq. \ref{eq:EQSshiftL2} that  $\gammal{h}{2}$ is associated with the frequency-shift of the $h$-th plasmonic mode.

The second order correction of the associated plasmonic mode $\Jl{h}{2}$ has both longitudinal and transverse components, denoted as $\Jll{h}{2}$ and $\Jlt{h}{2}$, respectively:
\begin{equation}
  \Jl{h}{2} =\Jll{h}{2} + \Jlt{h}{2}  = \sum_{\substack{k=1\\ k \ne h}}^\infty \ahk{h}{k}{2} \, \Jlo_{k} + \sum_{k=1}^\infty \beta_{h,k}^{\left( 2 \right)} \,  \Jto_k,
\end{equation}
where the longitudinal part $\Jll{h}{2}$ is represented in terms of the EQS  modes basis $\left\{ \Jlo_{k} \right\}_{k\in \mathbb{N}}$, and the transverse part $\Jlt{h}{2}$  in terms of the MQS  modes basis $\left\{ \Jto_{k} \right\}_{k \in \mathbb{N}}$. The expansion coefficients are \cite{SI}: 
\begin{widetext}
\begin{align}
\ahk{h}{k}{2} &=  \frac{1}{4\pi} \frac{\gammalo{k} \gammalo{h} }{\gammalo{k} - \gammalo{h}}  \left(  
      \oiint_{\partial \tOmega}   \Jlnn{h} \rp  \oiint_{\partial \tOmega}   \frac{\left| \rbt - \rbt' \right|}{ 2}\Jlnn{k} \left( \rbt' \right)
 \dS' \dS  +   \iiint_{\tOmega}  \Jlo_{h} \rp \cdot \iiint_{\tOmega} \frac{ \Jlo_{k} \rpp}{{\left| \rbt - \rbt' \right|}} \, \dV' \dV \right), \qquad \forall k \ne h
  \label{eq:alphal2}   \\
\beta_{h,k}^{\left( 2 \right)} &= \frac{1}{4\pi} \gammalo{h}   \iiint_{\tOmega}  \Jlo_{h} \rp \cdot \iiint_{\tOmega} \frac{ \Jto_{k} \rpp}{{\left| \rbt - \rbt' \right|}} \, \dV' \dV,  \,  \qquad \forall k \in \mathbb{N}.
\label{eq:betal2} 
\end{align}
\end{widetext}
Matching the third order terms in Eq. \ref{eq:VIEeqs}, the third order correction $\gammal{h}{3}$ is obtained: \cite{SI} 
\begin{equation}
   \gammal{h}{3}  = i \frac{1}{6 \pi} \left(\gammalo{h}\right)^2 \left\| \mathbf{P}_{\text{E},h}^\parallel \right\|^2,
   \label{eq:Chi3}
\end{equation}
which is purely imaginary and proportional to the squared norm of the dipole moment $\mathbf{P}_{\text{E},h}^\parallel $ of the $h$-th EQS mode.

As it will be demonstrated in  Eq. \ref{eq:Qlr3}, $ \gammal{h}{3}$ determines the radiative Q-factor of $h$-th plasmonic mode. However,  for dark plasmonic modes $\gammal{h}{3}$  vanishes. In this case, to retrieve information about the radiative Q-factor, it is mandatory to consider the fifth order perturbation $\gammal{h}{5}$.  For dark modes, it can be expressed \cite{SI} in terms of the electric quadrupole tensor $\tensor{\bf Q}^\parallel_{\text{E}|h}$ of the $h$-th EQS mode, and its components $Q_{\text{E} | h | ij}^\parallel$:

\begin{equation}
 \gammal{h}{5}  =   i \frac{1}{80\pi} \left(\gammalo{h}\right)^2  \left[ \sum_{ij}  \left( Q_{\text{E} | h | ij}^\parallel \right)^2   - \frac{1}{3} \left( \text{Tr} \, \tensor{\bf Q}^\parallel_{\text{E}} \right)^2 \right]
    \label{eq:Chi5}
 \end{equation}
where $\text{Tr}$ is the trace operator, and $\tensor{\bf Q}^\parallel_{\text{E}|h}$ is defined by Eq. \ref{eq:Quadrupole} of the Appendix \ref{sec:Multipoles}. Thus, the fifth order correction is purely imaginary  and proportional to the power radiated to infinity by the electric quadrupole $\tensor{\bf Q}^\parallel_{\text{E}}$. 

The outlined procedure can be iteratively applied: if the fifth order correction vanishes, the next order correction that may give an imaginary contribution is the seventh, which can be calculated by matching the terms of corresponding-order in Eq. \ref{eq:VIEeqs}.

\section{Dielectric Resonances}
\label{sec:dielectric}
To evaluate the dielectric resonances beyond the quasistatic regime it is convenient to recast the eigenvalue problem  \ref{eq:EigVIE} as
\begin{multline}
 - {x^2} \Jtoo \rp -  \kappa  \tnabla \oiint_{\partial \tilde{\Omega}} g \left( \rbt - \rbt', x \right) \Jtoo \rpp \cdot \n \rpp  \,  \dS'  \, \\ + \kappa \, {x^2} \iiint_{\tilde{\Omega}} g \left( \rbt - \rbt' , x \right) \Jtoo \rp \,  \dV' =  \mathbf{0} \quad \forall \rt\in \tilde{\Omega}.
 \label{eq:VIEt0}
\end{multline}
The  Green function $g \left( \rbt - \rbt' , x \right)$, the mode $\bf v$ and the corresponding eigenvalue $\kappa$ are expanded at $x=0$ in the neighborhood of MQS eigenvalue $\gammato{h}$ and mode $\Jto_{h}$:
\begin{align}
\label{eq:KappaExp}
\kappa_h &= \gammato{h} + \gammat{h}{1} x +  \gammat{h}{2} x^2+ \ldots = \sum_{k=0}^{\infty}  \gammat{h}{k} x^k, \\
\label{eq:JtExp}
\Jtoo_h &=  \Jto_{h} + \Jt{h}{1} x + \Jt{h}{2} \x{2} + \dots  =    \sum_{k=0}^{\infty} \Jt{h}{k} \x{k}.
\end{align} 
By substituting Eqs. \ref{eq:KappaExp}, \ref{eq:JtExp}, and \ref{eq:GExp} in Eq. \ref{eq:VIEt0} the following equation is obtained:
\begin{widetext}
\begin{multline}
 - {4 \pi} \sum_{k=0}^\infty \Jt{h}{k} \x{k+2}  -   \left( \sum_{k=0}^\infty \gammat{h}{k} \x{k} \right)  \tnabla  \oiint_{\partial \tOmega}  \left(  \sum_{k=0}^\infty (-i)^k \frac{\deltar^{k-1}}{k!} \x{k}  
    \right) \left( \sum_{k=0}^\infty \Jtn{h}{k} \x{k} \right) \dS' + \\ \left( \sum_{k=0}^\infty \gammat{h}{k} \x{k} \right) \iiint_{\tOmega}  \left(  \sum_{k=0}^\infty (-i)^k \frac{\deltar^{k-1}}{k!} \x{k}  
    \right) \left( \sum_{k=0}^\infty \Jt{h}{k} \x{k+2} \right) \dV' = 0, \qquad \forall \mathbf{r} \in \tOmega,
   \label{eq:VIEti} 
\end{multline}
\end{widetext}
where $\left. \Jtn{h}{k} = \Jt{h}{k} \cdot \mathbf{n}\right|_{\partial \Omega}$, $\gammat{h}{0} = \gammato{h}$ and $\Jt{h}{0}= \Jto_h$.
In the Supplemental Material  all the details on the derivation of radiation corrections for dielectric resonances are reported \cite{SI}. Here, only the main results are highlighted.  In the following, the MQS are normalized accordingly to Eq. \ref{eq:MQSorthonorm}, i.e. $\left\| \Jto_h \right\|=1$ $\forall h$.

By matching the terms of corresponding order in Eq. \ref{eq:VIEti}, it is possible to demonstrate that first order corrections vanish regardless of the shape of the object \cite{SI}:
\begin{align}
\gammat{h}{1} &= 0, \\
\Jt{h}{1} \left( \rbt \right) &= 0, \qquad \forall \mathbf{r} \in \tOmega.
\label{eq:Jt1n}
\end{align}

The second order correction $\gammat{h}{2}$ is a real quantity, namely \cite{SI}
\begin{multline}
\gammat{h}{2} =  \frac{ \left(\gammato{h}\right)^2}{4\pi} \left[ \iiint_{\tOmega}  \Jto_{h} \rp\cdot \iiint_{\tOmega}   \frac{\left| \rbt -\rbt'\right|}{2} \Jto_h \rpp \dV' \dV \right. \\ +  \left. \sum_{k=1}^\infty  \frac{\gammalo{k}}{4\pi}
  \left| \iiint_{\tOmega}  \Jlo_k  \rp \cdot  \iiint_{\tOmega}  \frac{ \Jto_{h} \rpp }{\left| \rbt -\rbt'\right|}   \dV' \dV \right|^2 \right]
   \label{eq:Kappa2b} 
\end{multline}
The first term in parenthesis in Eq. \ref{eq:Kappa2b} originates from the radiative self-interaction of the MQS mode $\Jto_{h}$ through the vector potential. The second terms is made of a summation, where each addend is proportional to the magnetostatic interaction energy between the  MQS current mode $\Jto_{h}$ and the EQS current mode $\Jlo_{k}$, denoted as $\mathscr{W}_{m \text{I}} \left\{ \Jlo_k ,\Jto_h \right\}$:
\begin{equation}
 \mathscr{W}_{m \text{I}} \left\{ \Jlo_k ,\Jto_h \right\} = \frac{\mu_0}{8 \pi} \iiint_{\tOmega} \Jlo_k \rp   \cdot \iiint_{\tOmega}  \frac{\Jto_h \rpp}{\left| \rbt -\rbt' \right|} \, \dV' \dV. 
\end{equation}

The MQS current mode $\Jto_{h}$ may be an $\mathbb{A}^\perp$-mode,  generating a  transverse vector potential, accordingly to the definition \ref{eq:Amode}. In this case, since every EQS current mode is longitudinal, and transverse and longitudinal functions are orthogonal accordingly to the scalar product \ref{eq:ScalarProd}, the energy $\mathscr{W}_{m \text{I}} \left\{ \Jlo_k ,\Jto_h \right\}$ vanishes $\forall k$, and Eq. \ref{eq:Kappa2b} further simplifies:
\begin{multline}
\gammat{h}{2} =  \frac{ \left(\gammato{h}\right)^2}{4\pi}   \iiint_{\tOmega}  \Jto_{h} \rp\cdot \iiint_{\tOmega}   \frac{\left| \rbt -\rbt'\right|}{2} \Jto_h \rpp \dV' \dV \\  \text{ for $\mathbb{A}^\perp$-modes}.
   \label{eq:Kappa2TE} 
\end{multline}
As it will be demonstrated in Eq. \ref{eq:MQSxres2}, $\gammat{h}{2} $ is associated with the frequency-shift of dielectric modes.

The second order correction $\Jt{h}{2}$ to the current density mode has both longitudinal and transverse components, denoted as $\Jtl{h}{2}$ and $\Jtt{h}{2}$, which can be in turn expanded in terms of EQS and MQS current modes, respectively:
\begin{equation}
\Jt{h}{2} = \Jtl{h}{2}  +  \Jtt{h}{2}   = \sum_{k=1}^\infty \alpha^{\left( 2 \right)}_{h,k} \, \Jlo_k +  \sum_{\substack{k=1 \\ k \ne h}}^\infty \beta_{h,k}^{\left( 2 \right)} \, \Jto_k
\label{eq:J2nExp}
\end{equation}
where
\begin{widetext}
\begin{align}
  \alpha^{\left( 2 \right)}_{h,k}  &= -  \frac{\gammalo{k}}{4 \pi} \iiint_{\tOmega} \Jlo_k \rp   \cdot \iiint_{\tOmega}  \frac{\Jto_h \rpp}{\left| \rbt -\rbt' \right|} \dV' \dV, \qquad \forall k, \\
 \beta_{h,k}^{\left( 2 \right)}  & = \frac{\gammato{k}\gammato{h} }{\gammato{h}-\gammato{k}} \frac{1}{4 \pi}   \left[   \iiint_{\tOmega}  \Jto_{k} \rp\cdot \iiint_{\tOmega}   \frac{\left| \rbt -\rbt'\right|}{2} \, \Jto_h \rpp \dV' \dV -  \sum_{s=1}^\infty   \alpha_{h,s}^{\left( 2 \right)}   \iiint_{\tOmega}  \Jlo_s  \rpp \cdot  \iiint_{\tOmega}  \frac{ \Jto_{k} \left( \rbt  \right)}{\left| \rbt -\rbt'\right|}   \dV' \dV    \right], \qquad \forall k \ne h.
\end{align}
\end{widetext}
Although any magnetoquasistatic mode has a zero electric dipole moment, its second order radiative correction $\Jt{h}{2}$ may exhibit a non-zero electric dipole moment $\mathbf{P}_{\text{E}|h}^{\left( 2 \right)} $, given by \cite{SI}
\begin{equation}
     \mathbf{P}_{\text{E}|h}^{\left( 2 \right)} =  \sum_{k=1}^\infty \alpha_{h,k}^{\left( 2 \right)} \,  \mathbf{P}_{\text{E}|k}^\parallel,
     \label{eq:DipoleMoment2}
\end{equation}
where $\mathbf{P}_{\text{E}|k}^\parallel$ is the electric dipole moment of the $k$-th EQS mode $\Jlo_k$. For $\mathbb{A}^\perp$-modes the longitudinal part of $\Jt{h}{2} $ vanishes, and they do not display electric dipole moment up to this order.

The third order correction $\gammat{h}{3}$ is purely imaginary and depends on the  magnetic dipole moment $\mathbf{P}_{\text{M},h}^\perp$ of the mode $\Jto_{h}$ \cite{SI}:

\begin{equation}
   \gammat{h}{3}  = i \left(\gammato{h}\right)^2  \frac{1}{6\pi} \left\| \mathbf{P}_{\text{M},h}^\perp \right\|^2,
   \label{eq:Kappa3}
\end{equation}
where $\mathbf{P}_{\text{M},h}^\perp$ is defined in Eq. \ref{eq:DipoleM}. The third order correction to the mode, i.e. $\Jt{h}{3}$, is purely transverse, thus can be expanded in the basis of MQS modes
\begin{equation}
 \Jt{h}{3} = \Jtt{h}{3} = \sum_{\substack{k=1 \\ k \ne h}}^\infty \beta_{h,k}^{\left( 3 \right)} \, \Jto_k
 \label{eq:Jtn3}
\end{equation}
where the expansion coefficient $\beta_{h,k}^{\left( 3 \right)}$  depends on the dot product of the magnetic dipole moments $\mathbf{P}_{\text{M}|h}^\perp$ and $\mathbf{P}_{\text{M}|k}^\perp $ of the $h$-th and $k$-th MQS mode \cite{SI}:
\begin{equation}
  \bhk{h}{k}{3}  = i   \frac{1}{6\pi}  \frac{\gammato{k}\gammato{h}}{\gammato{h} - \gammato{k}} \mathbf{P}_{\text{M}|h}^\perp \cdot \mathbf{P}_{\text{M}|k}^\perp  \qquad  \forall  h \ne k.
\end{equation}

As it will be shown in Eq. \ref{eq:QtrP}, the correction $\gammat{h}{3}$, if non-vanishing, determines the radiative Q-factor of the $h$-th dielectric mode. 

However, it vanishes when the corresponding magnetic dipole moment is zero. In this case, the next imaginary correction has order 5 and has the following expression:
\begin{multline}
\gammat{h}{5}  =  i 
   \frac{\left( \gammato{h} \right)^2}{80\pi}   \sum_{ij}  \left( Q_{\text{M}|h|ij}^\perp \right)^2   + \\  i    \frac{\left( \gammato{h} \right)^2}{6\pi} \left\|    \mathbf{P}_{\text{E}2|h}^\perp - \mathbf{P}_{\text{E}|h}^{\left( 2 \right)}   \right\|^2 
\label{eq:kappa5b}
\end{multline}
where $\tensor{\mathbf{Q}}_{\text{M}|h}^\perp$ is the magnetic quadrupole tensor of the $h$-th MQS mode, introduced in Eq. \ref{eq:QuadrupoleM} and $ Q_{\text{M}|h|ij}^\perp $ are its components, $\mathbf{P}_{\text{E}2|h}^\perp$ is the toroidal dipole defined in Eq. \ref{eq:ToroidalM}, and  $\mathbf{P}_{\text{E}|h}^{\left( 2 \right)}$ is the electric dipole moment of the second order correction $\Jt{h}{2}$, introduced in Eq. \ref{eq:DipoleMoment2}. {\color{red} In conclusion, the fifth order correction is determined by two contributions: they accounts for the power radiated to infinity by the magnetic quadrupole $\tensor{\mathbf{Q}}_{\text{M}|h}^\perp$, and by an effective electric dipole resulting from the interference between the $\mathbf{P}_{\text{E2}|h}^\perp$ and $\mathbf{P}_{\text{E}|h}^{\left( 2 \right)}$.

For  $\mathbb{A}^\perp$-modes Eq. \ref{eq:kappa5b} further simplifies
\begin{equation}
\gammat{h}{5}  = i \left( \gammato{h} \right)^2  \left\{ 
   \frac{1}{80\pi}   \displaystyle\sum_{ij}  \left( Q_{\text{M}|h|ij}^\perp \right)^2     +  \frac{1}{6\pi} \left\|    \mathbf{P}_{\text{E}2|h}  \right\|^2   \right\}
   \label{eq:kappa5TE}
 \end{equation}
}

The outlined procedure can be iteratively applied. If the fifth order correction vanishes, the next order correction that may give an imaginary contribution is the seventh, which can be calculated by matching the terms of nine-th order in Eq. \ref{eq:VIEti}.

\section{Resonance Frequency and Q-factor}
\label{sec:Q}
In the previous section, the second order corrections $\gammal{h}{2}$, $\gammat{h}{2}$ and non-vanishing imaginary corrections  $\gammal{h}{n_i}$, $\gammat{h}{n_i}$ of the lowest order, called $n_i$, are derived in closed form for both plasmonic and dielectric modes. They  depend neither on the size of the object nor on its permittivity, but they only depend on the morphology of the EQS and MQS modes. In this section, closed form expressions of the resonance frequency and Q-factors  are obtained in terms of these corrections for both metal and dielectric objects. The modes are assumed to be non-interacting.  Moreover, throughout this work we use the definition of Q-factor as the inverse of the $-3$dB fractional bandwidth.

\subsection{Plasmonic Resonances}
It is now assumed that the object is made of a time-dispersive metal described by the Drude model \cite{kreibig2013optical,maier2007plasmonics}
\begin{equation}
  \chi \left( \omega \right) = - \frac{\omega_p^2}{\omega \left( \omega - i \nu \right)},
  \label{eq:EQSxres0}
\end{equation}
where $\omega_p$ and $\nu$ are the plasma and collision angular frequencies, and $\nu \ll \omega_p$. It is also useful to define the quantity
\begin{equation}
x_p =  \frac{\omega_p}{c_0} \, l_c = 2 \pi \frac{l_c}{\lambda_p},
\end{equation}
where $\lambda_p$ is the plasma wavelength. The EQS resonance frequency $\omega_h^\parallel$ of the $h$-th mode, is defined as the frequency at which the real part of the metal susceptibility  $ \re{\chi \left( \omega \right)}$  matches the EQS eigenvalue $\gammalo{h}$, i.e.   
\begin{equation}
   \frac{x_h^\parallel}{x_p} = \frac{\omega_h^\parallel}{\omega_p} = \frac{1}{\sqrt{-\gammalo{h}}},
\end{equation}
where $x_h^\parallel = \left( \omega_h^\parallel/c_0 \right) l_c$ is the size parameter at the EQS resonance.

In the full-wave scenario, the resonance of the $h$-th plasmonic mode is defined by 
setting to zero the real part of the denominator of the $h$-th addend of the first summation in Eq. \ref{eq:MIMexpansion}. Thus, the value $x_h$ of the size-parameter at the plasmonic resonance is the value of $x$ at which the real part of the metal susceptibility  $\chi \left( \omega \right)$  matches the real part of the corresponding eigenvalue $\chi_h \left( x_h \right)$ of Eq. \ref{eq:EigVIE}. i.e. 
\begin{equation}
 \re{\chi_h} =   \re{\chi \left( \omega_h \right)} \approx -\frac{\omega_p^2}{\omega_h^2} = - \frac{x_p^2}{x_h^2} ,
   \label{eq:ResCondPlasmon}
\end{equation}
and $\omega_h$ is the corresponding resonance frequency. Eq.  \ref{eq:ResCondPlasmon}  is the resonance condition of the plasmonic modes. For small particles $x_p \lessapprox 1$,  by retaining only the  real and imaginary non-zero corrections of the lowest order in Eq. \ref{eq:ChiExp},
the plasmonic eigenvalue $ \chi_h \left( x \right)$ is approximated as
\begin{equation}
 \chi_h \left( x \right)  \approx  \gammalo{h} + \gammal{h}{2} x^2 +  \gammal{h}{\nim} x^{\nim},
 \label{eq:ChiExp2}
\end{equation}
where $\nim$ is the order of the first non-zero imaginary correction $\gammal{h}{n_i}$. 
By using Eq. \ref{eq:ChiExp2} in \ref{eq:ResCondPlasmon}, and solving the resulting biquadratic equation, the resonance frequency is obtained: 
\begin{equation}
\frac{\omega_h}{\omega_p} =  \frac{x_h}{x_p}  = \frac{1}{\sqrt{2} x_p} \sqrt{ \frac{\gammalo{h}}{\gammal{h}{2}} \left( \sqrt{1 - 4  \frac{\gammal{h}{2}}{\left( \gammalo{h} \right)^2} \, x_p^2 } -1 \right)}.
\label{eq:EQSxres2}
\end{equation}
In the limit $x_p \ll 1 $, the frequency shift of the plasmonic resonance  with respect to the EQS resonance, i.e. $\Delta \omega_h = \omega_h -\omega_h^\parallel$, and the corresponding shift in the resonance size parameter, i.e. $\Delta x_h = x_h - x_h^\parallel$, can be approximated as
\begin{equation}
   \frac{\Delta \omega_h}{\omega_h^\parallel} = \frac{\Delta x_h}{x_h^\parallel}  \approx - \frac{1}{2} \frac{\gammal{h}{2}}{{\gammalo{h}}} \left( x_h^\parallel \right)^2, \qquad x_p \ll 1.
   \label{eq:EQSshiftL2}
\end{equation}
In conclusion, the relative  frequency shift of any plasmonic mode is a quadratic function of $x_h^\parallel $, whose prefactor is one half the ratio between the second order correction $\gammal{h}{2}$ and the EQS eigenvalue $\gammalo{h}$.

 \begin{figure*}[!ht]
\centering
\includegraphics[width=\textwidth]{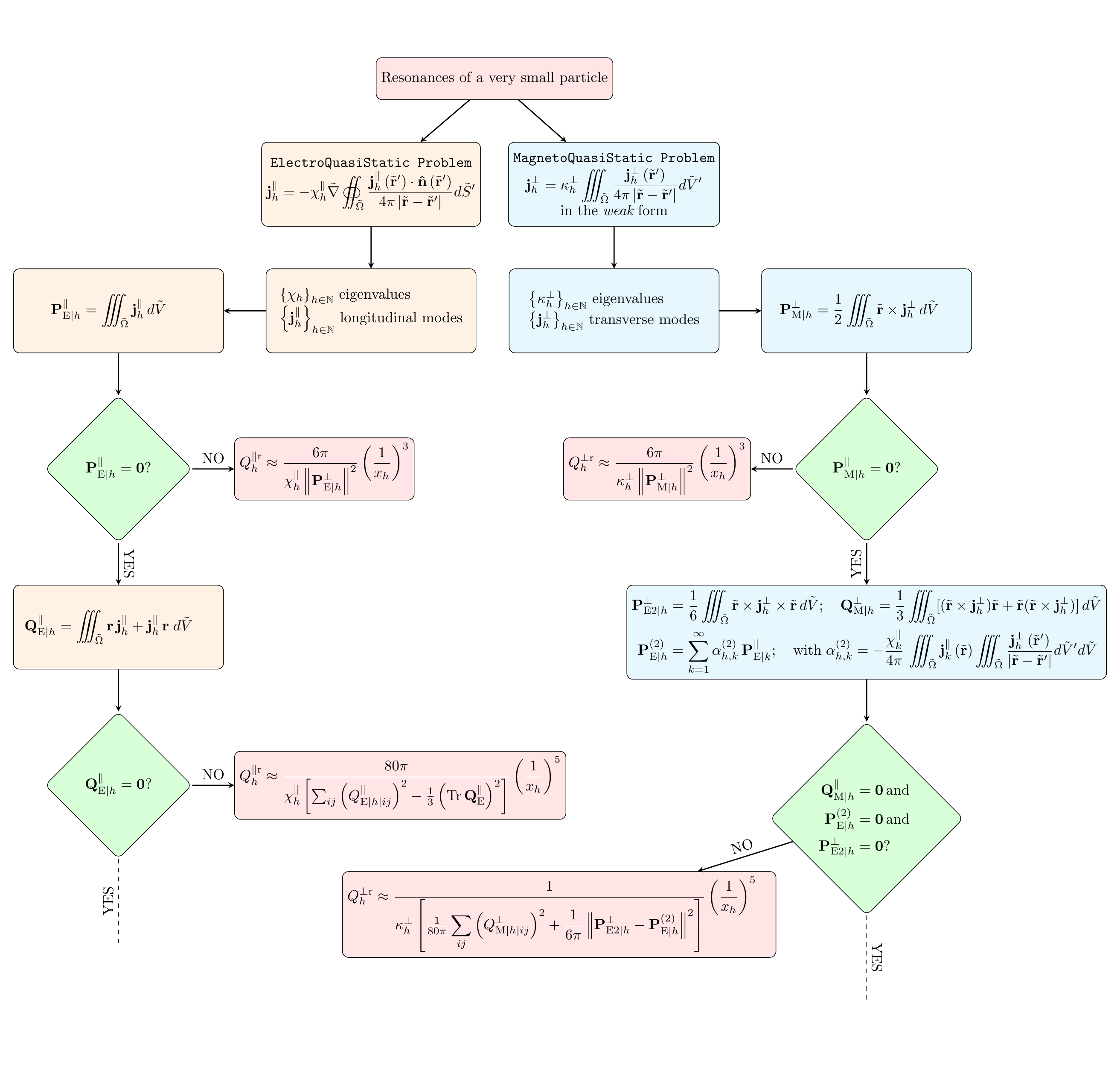}
\caption{Algorithm to compute the radiative Q-factor of either plasmonic or dielectric resonances of a small homogeneous isotropic non-magnetic object. In the plasmonic case, the EQS current modes and corresponding eigenvalues are computed. If the $h$-th EQS mode is bright with non-vanishing electric dipole moment $\PE{h}$, its radiative Q-factor is given by Eq. \ref{eq:Qlr3}. Instead, for a dark mode with non vanishing electric quadrupole tensor $\QE{h}$, the Q-factor given by  \ref{eq:Qlr5}. If $\QE{h}$ is also vanishing, the outlined process can be iterated by considering higher order electric multipoles. In the dielectric case, the MQS current modes and corresponding eigenvalues are computed. Thus, if the $h$-th MQS mode exhibits a non-vanishing magnetic dipole moment $\PM{h}$, the radiative Q-factor is given by Eq. \ref{eq:QtrP}. Instead, if the magnetic dipole vanishes, but at least one among the magnetic quadrupole  $\QM{h}$, the toroidal dipole  $\TE{h}$, or the electric dipole moment of the second order mode correction $\mathbf{P}_{\text{E}|h}^{\left( 2 \right)}$ is non-vanishing, the radiative Q-factor is given by Eq. \ref{eq:QtrQ}. If they are all vanishing, the outlined algorithm has to be iterated, and higher order multipoles considered.}
\label{fig:FlowChart}
\end{figure*}

The radiative Q-factor $\Qlr{h}$ of the $h$-th plasmonic mode is obtained by considering the inverse of the fractional bandwidth of the $h$-th addend of the first summation in Eq. \ref{eq:MIMexpansion}, assuming negligible non-radiative losses, and  using the expansion \ref{eq:ChiExp2}:
\begin{equation}
 \label{eq:eqsQr}
  \Qlr{h}   \approx  \left|\frac{ \gammalo{h} }{\gammal{h}{\nim}}\right| \left( \frac{1}{x_h} \right)^{\nim}.
\end{equation}
The Q-factor is an inverse power function of the size parameter at the resonance, whose exponent is the order $n_i$ of the first non-vanishing imaginary correction $\gammal{h}{\nim}$, while the prefactor is the ratio between the EQS eigenvalue $\gammalo{h}$ and the correction $\gammal{h}{\nim}$.

In Fig. \ref{fig:FlowChart}, the algorithm for the calculation of the radiative Q-factor of any plasmonic mode is summarized by a flowchart. First, the EQS current modes and corresponding eigenvalues are computed by solving the eigenvalue problem \ref{eq:EQSproblem}. Thus, if the $h$-th mode is bright, namely it exhibits a non-vanishing electric dipole moment $\mathbf{P}_{\text{E}|h}^\parallel$, it follows that $\gammal{h}{3} \ne 0$, $n_i=3$, and the Q-factor is obtained by combining Eqs. \ref{eq:Chi3} and \ref{eq:eqsQr}:
\begin{equation}
  \Qlr{h}  \approx  \frac{6\pi}{  \gammalo{h}  \left\| \mathbf{P}_{\text{E}|h}^\parallel \right\|^2} \left( \frac{1}{x_h} \right)^{3}.
  \label{eq:Qlr3}
\end{equation}
Instead, for a dark mode with non vanishing electric quadrupole tensor,  it follows that $\gammal{h}{5} \ne 0$, $n_i=5$, and the Q-factor is determined by combining Eqs. \ref{eq:Chi5} and \ref{eq:eqsQr}:
\begin{equation}
 \Qlr{h} \approx
\frac{80\pi}{  \gammalo{h} \left[ \sum_{ij}  \left( Q_{\text{E} | h | ij}^\parallel \right)^2   - \frac{1}{3} \left( \text{Tr} \, \tensor{\bf Q}^\parallel_{\text{E}|h} \right)^2 \right] } \left( \frac{1}{x_h} \right)^{5}.
  \label{eq:Qlr5}
 \end{equation}

If the electric quadrupole moment is also vanishing, the outlined process can be iterated by considering higher order electric multipoles.

For completeness, we also consider the opposite regime, dominated by non-radiative losses. In this case, the non-radiative Q-factor $\Qld{h}$ is obtained as the inverse of the fractional bandwidth of the $h$-th addend of the first summation in Eq. \ref{eq:MIMexpansion}, assuming negligible radiative losses. It has the expression:
\begin{equation}
 \label{eq:eqsQnr}
 \Qld{h} = \frac{\omega_h}{\nu}  \approx \frac{\omega_p}{ \nu \sqrt{-\chi}}.
\end{equation}
Equation \ref{eq:eqsQnr} is not new, but it was already shown by Wang and Shen in Ref. \cite{Wang:06}.

In an intermediate regime, the resulting Q-factor, indicated with $\Qlt{h}$, can be obtained as \cite{johnson1939transmission}
\begin{equation}
  \frac{1}{\Qlt{h}} = \frac{1}{\Qlr{h}} + \frac{1}{\Qld{h}}.
   \label{eq:eqsQt}
\end{equation}


\subsection{Dielectric Resonances}
It is now assumed that the object is made of a non-dispersive dielectric material with positive susceptibility $\chi \ge 0$, with $\im{\chi} \ll \re{\chi}$.  The size parameter $x_h^\perp =  \left( \omega_h^\perp/c_0 \right) l_c$ at the MQS resonance  is defined as the value of $x$ at which the real part of the susceptibility $\chi$ matches the  eigenvalue $\gammato{h}/x^2$, namely: 
\begin{equation}
  x_h^{\perp} = \frac{\omega_h^\perp}{c_0} l_c = \sqrt{\frac{ \gammato{h}}{\re{\chi}}},
  \label{eq:MQSxres0}
\end{equation}
and $\omega_h^\perp$ is the corresponding MQS resonance frequency.

In the full-wave regime, the resonance of the $h$-th dielectric mode is defined by 
setting to zero the real part of the denominator of the $h$-th addend of the second summation in Eq. \ref{eq:MIMexpansion}. Thus, the value of size parameter $x_h = \left( \omega_h/c_0 \right) l_c$  at the dielectric resonance is the value of $x$  at which the real part of the eigenvalue $\kappa_h/x^2$ matches the quantity $\re{\chi}$
\begin{equation}
   \re{\kappa_h} = \re{\chi} x^2.
    \label{eq:ResonanceConditionMQS}
\end{equation}
This is the resonance condition for dielectric modes, and $\omega_h$ is the dielectric resonance frequency. For small particles $x \lessapprox 1$,  by keeping only the real and imaginary non-zero corrections of the lowest order in Eq. \ref{eq:KappaExp}, the dielectric eigenvalue $\kappa_h \left( x \right)$ is approximated as:
\begin{equation}
\kappa \left( x \right) \approx  \gammato{h} + \gammat{h}{2} {x}^2 +  \gammat{h}{\nim} {x}^{\nim},
\label{eq:KappaExp2}
\end{equation}
where $\nim$ is the  order of the first non-zero imaginary correction $\gammat{h}{\nim} $.
By using Eq. \ref{eq:KappaExp2} in \ref{eq:ResonanceConditionMQS}, and solving the resulting quadratic equation, the resonance size parameter is obtained:
\begin{equation}
  x_h = \frac{\omega_h}{c_0}{l_c} \approx \sqrt{ \frac{\gammato{h}}{\re{\chi} - \gammat{h}{2}}} =      \frac{x_h^{\perp}}{\sqrt{1- {\gammat{h}{2}}/{\re{\chi}}}}.
  \label{eq:MQSxres2}
\end{equation}
For high-index dielectrics $ \re{\chi}  \gg 1$, the relative frequency  shift of the $h$-th dielectric resonance with respect to the MQS resonance frequency is
\begin{equation}
 \frac{\Delta \omega_h}{\omega_h^{\perp}}  =  \frac{\Delta x_h}{x_h^{\perp}} \approx \frac{\gammat{h}{2}}{\re{\chi}} =  \frac{\gammat{h}{2}}{\gammato{h}}  \left( x_h^{\perp} \right)^2, \qquad  \re{\chi}  \gg 1,
 \label{eq:MQSshift}
\end{equation}
In conclusion, the relative  frequency-shift of any dielectric mode is a quadratic function of $x_h^{\perp}$, whose prefactor is approximately the ratio between the second order correction $\gammat{h}{2}$ and the quasistatic eigenvalue $\gammato{h}$.

The radiative Q-factor $\Qtr{h}$ of the $h$-th dielectric mode is obtained as the inverse of the fractional bandwidth of the $h$-th addend of the second summation in Eq. \ref{eq:MIMexpansion}, assuming negligible non-radiative losses $\im{\chi}\approx 0$, and using the expansion \ref{eq:KappaExp2}:
\begin{equation}
  \Qtr{h} \approx \left| \frac{ \gammato{h} }{\gammat{h}{\nim}} \right| \left( \frac{1}{x_h} \right)^{\nim}.
    \label{eq:mqsQr}
 \end{equation}
The  radiative Q-factor is an inverse power function of the size parameter whose exponent is the order $n_i$ of the first non-vanishing imaginary correction $\gammat{h}{\nim}$, while the prefactor is the ratio between the quasistatic eigenvalue $\gammato{h}$ and $\gammat{h}{\nim}$. 

In Fig. \ref{fig:FlowChart}, the algorithm for the calculation of the radiative Q-factor of any dielectric mode is summarized by a flowchart. First, the MQS current modes and corresponding eigenvalues are computed by solving the eigenvalue problem \ref{eq:MQSproblem}. Thus, if the mode exhibits a non-vanishing magnetic dipole moment, $ \Qtr{h} $ has the following expression, obtained by combining Eq. \ref{eq:mqsQr} and \ref{eq:Kappa3}:
\begin{equation}
  \Qtr{h} \approx  \frac{6\pi}{  \gammato{h}  \left\| \mathbf{P}_{\text{M}|h}^\perp \right\|^2} \left( \frac{1}{x_h} \right)^{3}.
  \label{eq:QtrP}
\end{equation}
Instead, if the magnetic dipole vanishes, but at least one among the magnetic quadrupole moment $\QM{h}$, the toroidal dipole moment $\TE{h}$, or the dipole moment of the second order mode correction $\mathbf{P}_{\text{E}|h}^{\left( 2 \right)}$ is non-vanishing, the radiative Q-factor has the following expression
\begin{widetext}
\begin{equation}
  \Qtr{h}  \approx \frac{1}{ \gammato{h}  \left[
  {\frac{1}{{80\pi}}}   \displaystyle\sum_{ij}  \left( Q_{\text{M}|h|ij}^\perp \right)^2     +  \frac{1}{6\pi} \left\|    \mathbf{P}_{\text{E}2|h}^\perp -  \mathbf{P}_{\text{E}|h}^{\left( 2 \right)} \right\|^2    \right] } \left( \frac{1}{x_h} \right)^{5}.
\label{eq:QtrQ}
\end{equation}
 \end{widetext}
If they are all vanishing, the outlined algorithm has to be iterated, and higher order multipoles have to be considered.
  
In dielectric resonators  the opposite regime, dominated by non-radiative losses,  is less common. Nevertheless, it is now considered for completeness. In this case, the non-radiative Q-factor $\Qtd{h}$ is obtained as the inverse of the factional bandwidth of the $h$-th addend of the second summation in Eq. \ref{eq:MIMexpansion}, assuming dominating non-radiative losses:
\begin{equation}
 \Qtd{h} = \frac{ \left(\gammato{h}\right)^2 }{\text{Im}\left\{ \chi \right\}} \left( \frac{1}{x_h} \right)^2 \approx \frac{ \text{Re} \left\{ \chi \right\}}{ \text{Im} \left\{ \chi \right\}},
  \label{eq:mqsQnr}
\end{equation}
In an intermediate regime, the Q-factor, indicated with $\Qtt{h}$,  is obtained as\cite{johnson1939transmission}

\begin{equation}
  \frac{1}{\Qtt{h}} = \frac{1}{\Qtr{h}} + \frac{1}{\Qtd{h}}.
  \label{eq:mqsQt}
\end{equation}

\section{Results and Discussion}
\label{sec:Tables}
Once the shape of an homogeneous object is assigned,  its {\it catalogues} of plasmonic and dielectric modes can be introduced. The two catalogues  are shown in Figs. \ref{fig:ModesS_L} and \ref{fig:ModesS_T} for a sphere, in Figs.  \ref{fig:ModesC_L} and  \ref{fig:ModesC_T} for a finite-size cylinder, and in Figs. \ref{fig:ModesT_L} and  \ref{fig:ModesT_T} for a triangular prism.  The catalogues illustrate the EQS and MQS current modes,  where each set is ordered according to their real quasistatic eigenvalue $\gammalo{h}$ and $\gammato{h}$, respectively.  Besides the quasistatic eigenvalue, each resonance is also characterized by the second order correction $\gammat{h}{2}$ and $\gammal{h}{2}$, and by the lowest-order (non-vanishing) imaginary correction $\gammat{h}{\nim}$, $\gammal{h}{\nim}$, where $\nim$ is odd with $\nim \ge 3$.  In both tables, the value of $\nim$ is also highlighted, enclosed in a circle on the top-right of each box.  Plasmonic current modes  are also labeled with a ``$\mathbb{D}$'' if   dark, while dielectric  modes are labeled with ``$\mathbb{A}^\perp$'' if they generate a transverse vector potential, accordingly to the definition \ref{eq:Amode}.  The information contained in these two catalogues depends neither on the permittivity of the object nor on its size nor on the frequency of operation; it only depends  on the morphology of the object.  

The tables of plasmonic and dielectric resonances contain essential information to characterize and engineer the electromagnetic scattering of small objects. Specifically, the relative frequency shift of both plasmonic and dielectric resonances is a quadratic function of the size parameter at the quasistatic resonance, whose prefactor is $- \gammal{h}{2}/ 	\left( 2 \, \gammalo{h} \right)$ for plasmonic modes and $\gammat{h}{2}/\gammato{h}$ for dielectric resonances, respectively. Furthermore, the {\it radiative} Q-factor is an inverse power function of the size parameter  at the resonance, whose exponent is exactly the order $\small{\nim}$, while the prefactor is the ratio $\gammalo{h}/\gammal{h}{\nim}$ for plasmonic and $\gammato{h}/\gammat{h}{\nim}$ for dielectric modes. This prefactor  only  depends  on the quasistatic eigenvalue and on the multipolar components of the quasistatic mode.
\label{sec:Results}

In this manuscript, the electrostatic eigenvalue problem \ref{eq:EQSproblem} is solved by the surface integral method outlined in Refs. \cite{Fredkin:03,Mayergoyz:05} using a triangular mesh. The magnetoquasistatic eigenvalue problem \ref{eq:MQSproblem} is solved by the numerical method outlined in Refs. \cite{Forestiere:20} by using loop basis functions defined on a hexahedral mesh. Then, the radiation correction for both plasmonic (Eqs. \ref{eq:Chi2b}, \ref{eq:Chi3}, and \ref{eq:Chi5}) and dielectric eigenvalues (Eqs. \ref{eq:Kappa2b}, \ref{eq:Kappa3}, and \ref{eq:kappa5b}) are computed using standard quadrature formulas, and, if singular, using the formulas provided by R. Graglia \cite{Graglia:87,graglia1993numerical}.

\subsection{Sphere}

The plasmonic and dielectric resonances of a sphere of radius $R$ are now investigated. The characteristic length $l_c$ is assumed equal to the radius $R$. The sphere is the ideal shape to numerically validate the radiation corrections, because the quasistatic modes, the corresponding eigenvalues, and radiation corrections have analytic expression, which is given in appendix \ref{sec:AnalyticSphere}. In the current section, these same quantities are  calculated numerically and compared to their analytic counterparts. In particular, the surface mesh used for the calculation of EQS modes has $1500$ nodes, and $2996$ triangles, while the hexahedral mesh used for the calculation of MQS modes has $6527$ nodes, $6048$ hexahedra, and $11665$ edges. The same two meshes are also used for the computation of the surface and volume integrals of the radiation corrections. In the Supplemental Material \cite{SI}, the radiation corrections of Eqs. \ref{eq:Chi2b}, \ref{eq:Chi3} and of \ref{eq:Kappa2b}, \ref{eq:Kappa3} are analytically calculated in few scenarios.

\subsubsection{Catalogue of plasmonic resonances}

\begin{figure*}[!ht]
\centering
\includegraphics[width=\textwidth]{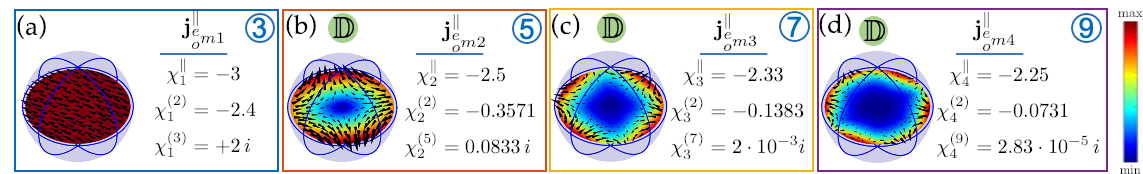}
\caption{Catalogue of plasmonic resonances of a sphere. The electroquasistatic current density modes are ordered according to their eigenvalues $\gammalo{n}$. The second order correction $\gammal{n}{2}$ and the non-vanishing imaginary correction $\gammal{n}{n_i}$ of lowest-order $n_i$ are shown on the right of the corresponding box, while the value of $n_i$ is shown on the top-right  enclosed in a circle. The {\it dark} modes are also labeled with ``$\mathbb{D}$''.}
\label{fig:ModesS_L}
\end{figure*}

The {\it catalogue}  of plasmonic resonances of a sphere is shown in Fig. \ref{fig:ModesS_L}. The radiation corrections   are calculated numerically and analytically, and the numerical error is shown in Table \ref{tab:EQS}. The three degenerate EQS modes $\Jlo_{\substack{e \\ o}m1}$ with $m=0,1$ (namely $\Jlo_{e01}$, $\Jlo_{e11}$, and $\Jlo_{o11}$) are associated to the lowest eigenvalue $\chi_1^\parallel=-3$.  The analytic expression of the modes is given in Eq. \ref{eq:EQSmode} of the appendix \ref{sec:AnalyticSphere}, while one of them is depicted in Fig. \ref{fig:ModesS_L} (a). They are bright and represent three electric dipoles oriented along mutually orthogonal directions. Their second order correction $\gammal{1}{2} = -2.4$ is calculated both numerically, performing the integrals in Eq. \ref{eq:Chi2b}, and analytically by using Eq. \ref{eq:EQSsphere2}. The third order correction $\gammal{1}{3} = +2 \, i$ is proportional to the squared magnitude of their electric dipole moment, and it is calculated both numerically by Eq. \ref{eq:Chi3} and analytically by Eq. \ref{eq:EQSsphere3}.

\begin{table}[h!]
\centering
\begin{tabular}{c|c|c|c}
\hline
$\#$ &  &   $\Jlo_{\substack{e \\ o}m1}$ &  $\Jlo_{\substack{e \\ o}m2}$   \\
\hline
\hline
\multirow{3}{*}{$\gammalo{n}  $}  & numeric  & -3.00 & -2.50   \\   
 &  analytic  & -3  & -2.5   \\
 &  error $ \left[ \% \right]$ & 0.12 & 0.2    \\
\hline
\hline
  \multirow{3}{*}{$\gammal{n}{2}  $}  & numeric & -2.38 &  -0.350    \\
  & analytic & -2.4 &    -0.357    \\
  & error $ \left[ \% \right]$ & 1.1  & 3.0    \\
\hline
\hline
  \multirow{3}{*}{$\gammal{n}{3}  $}  & numeric &  1.98  & $ \sim 10^{-9}$    \\
   &  analytic  &  2 & 0     \\
  & error $ \left[ \% \right]$ &  1.17   & -  \\
  \hline
  \hline
  \multirow{3}{*}{$\gammal{n}{5}  $} & numeric & -  & 0.0833 \\
   &  analytic  & - & 0.0816  \\
   & error $ \left[ \% \right]$ & -  & 2.1    \\
  \hline
\end{tabular}
\caption{EQS eigenvalues of a sphere and their radiative corrections, obtained numerically by Eqs. \ref{eq:Chi2b}, \ref{eq:Chi3}, \ref{eq:Chi5}, and analytically by Eqs. \ref{eq:EQSsphere}. Relative error.}
\label{tab:EQS}
\end{table}

The next five degenerate EQS modes are $\Jlo_{\substack{e \\ o}m2}$ with $m=0,1,3$. They are dark, and exhibit a non-vanishing electric quadrupole tensor. These modes are exemplified in Fig. \ref{fig:ModesS_L} (b). The second order correction is $\gammal{2}{2} = -0.35$. {\color{red} The third order correction $\gammal{2}{3}$ vanishes since these modes are dark.} The imaginary correction of the lowest order is $\gammal{2}{5}=0.0833i$, given by Eq. \ref{eq:Chi5} and analytically by Eq. \ref{eq:EQSsphere3}. It only depends on the electric quadrupole tensor of the EQS mode.

A similar line of reasoning  can be also applied to octupolar modes  $\Jlo_{\substack{e \\ o}m3}$ shown in  Fig.  \ref{fig:ModesS_T} (c), and to the hexadecapolar modes $\Jlo_{\substack{e \\ o}m4}$ shown in  Fig.  \ref{fig:ModesS_T} (d).

\subsubsection{Catalogue of dielectric resonances}

\begin{figure*}[!ht]
\centering
\includegraphics[width=\textwidth]{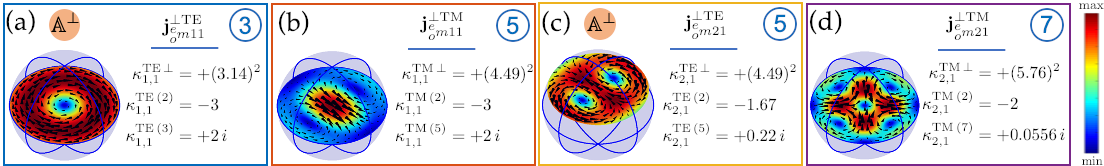}
\caption{Catalogue of dielectric resonances of a sphere. The magnetoquasistatic current density modes are ordered according to their eigenvalue. The second order correction, and the non-vanishing imaginary correction of lowest-order $n_i$ are shown on the right of the corresponding box, while the value of $n_i$ of each mode is shown on the top-right enclosed in a circle. The current modes generating a transverse vector potential are labeled with ``$\mathbb{A}^\perp$''. }
\label{fig:ModesS_T}
\end{figure*}

The {\it catalogue} of dielectric resonances of a sphere is shown in Fig. \ref{fig:ModesS_T}. The radiation corrections are calculated numerically and analytically, and the numerical error is shown in Table \ref{tab:MQS}.  The MQS current modes $\jtTE{m11}$ with $m=0,1$ (namely $\mathbf{j}_{e011}^{\text{TE}\perp}$, $\mathbf{j}_{e111}^{\text{TE}\perp}$, and $\mathbf{j}_{o111}^{\text{TE}\perp}$) are associated to the lowest MQS eigenvalue, i.e. $\kappa^{\text{TE} \, \perp}_{11}$.  The  analytic expressions of the modes and of the corresponding MQS eigenvalues are given in Eq. \ref{eq:MQSmodeTE}, and Eq. \ref{eq:MQSte0} of Appendix \ref{sec:AnalyticSphere}. The current modes $\jtTE{m11}$  are three degenerate magnetic dipoles oriented along three orthogonal axis; one of them is shown in Fig. \ref{fig:ModesS_T} (a). They are $\mathbb{A}^\perp$-modes because they generate a transverse vector potential, namely with zero normal component to $\partial \tOmega$, accordingly to the definition given in Sec.  \ref{sec:MQS}. Thus, the second order correction $\kappaTE{\left(2\right)}{11}$ has the simplified expression  \ref{eq:Kappa2TE}, because the magnetostatic interaction energy between $\jtTE{m11}$ and any EQS current mode vanishes, i.e. $\mathscr{W}_{m \text{I}} \left\{ \Jlo_{\substack{e \\ o}m'n'} , \jtTE{m11} \right\} = 0 $ $\forall m',n'\ge m'$.  The second order correction also features the analytic expression \ref{eq:MQSte2}. The third order correction $\kappaTE{\left(3\right)}{11}$ is given by Eq. \ref{eq:Kappa3} and is proportional to the squared magnitude of the magnetic dipole moment of the mode;  its analytic expression is given in Eq. \ref{eq:MQSte3}.

The next three degenerate modes, namely $\jtTM{m11}$ with $m=0,1$, are shown in Fig. \ref{fig:ModesS_T} (b). The mode analytic expression is given in Eq. \ref{eq:MQSmodeTM}. Each of them generates a vector potential with a non-vanishing longitudinal component. Specifically, the magnetostatic interaction energies between the current modes $\jtTM{m11}$  and the EQS modes $\Jlo_{\substack{e \\ o}m'1}$ with $m'=0,1$ (shown in Fig. \ref{fig:ModesS_L} (a)) is non-vanishing. They contribute to the second order correction $\kappaTM{\left(2\right)}{11}$, as prescribed by Eq. \ref{eq:Kappa2b}. The second order correction has also the analytic expression given in Eq. \ref{eq:MQStem2}. The third order correction $\kappaTM{\left(3\right)}{11}$ vanishes because the magnetic dipole moment of these modes is zero. Thus, the first non vanishing imaginary correction is $\kappaTM{\left(5\right)}{11}$, given by Eq. \ref{eq:kappa5b}, and analytically by Eq. \ref{eq:MQStem3}.  Since  $\jtTM{m11}$ have a vanishing magnetic quadrupole tensor, the correction $\kappaTM{\left(5\right)}{11}$  originates only from an effective dipole moment resulting from the interplay between the dipole moment of the second order mode correction and the toroidal dipole.

The next five degenerate modes are $\jtTE{m21}$ with $m=0,1,2$ and are shown in Fig. \ref{fig:ModesS_T} (c). As already pointed out in Ref. \cite{Forestiere:20} the modes  $\jtTE{m21}$ have the same MQS eigenvalue of   $\jtTM{m11}$. Nevertheless, differently from them, they are $\mathbb{A}^\perp$-modes.  For this reason the second order correction $\kappaTE{\left(2\right)}{21}$ has the simplified expression  \ref{eq:Kappa2TE}. Their magnetic dipole moment is zero, thus the third order correction vanishes.  The electric toroidal dipole moment is zero as well as the dipole moment of the second order mode correction. Nevertheless, the magnetic quadrupole tensor is non-vanishing and the fifth order correction $\kappaTE{\left(5\right)}{21}$  can be calculated by Eq. \ref{eq:kappa5TE}.

Similar line of reasoning can be also applied to the degenerate modes $\jtTE{m21}$ with $m=0,1,2$, shown in  Fig. \ref{fig:ModesS_T} (d).


\begin{table}[h!]
\centering
\begin{tabular}{c|c|c|c|c}
\hline
&  & $\jtTE{m11}$& $\jtTM{m11}$ & $\jtTE{m21}$\\
\hline
\multirow{3}{*}{$\sqrt{\gammato{nl}}$} & numeric  & 3.16  & 4.53  & 4.52   \\   
 & analytic  & 3.14  & 4.49 & 4.49 \\
 & error $ \left[ \% \right]$ & 0.64  & 1.81 & 1.04 \\
  \hline
  \hline
  \multirow{3}{*}{$\gammat{nl}{2}$} & numeric & -3.02  & -3.08 & -1.69 \\
   & analytic & -3 &  -3 & -1.67  \\
   & error $  \left[ \% \right]$ & 0.67 & 2.3 &  1.2  \\
  \hline
  \hline
  \multirow{3}{*}{$\gammat{nl}{3}$} & numeric  & 2.0 & 0 &  0  \\
  & analytic &  2 & 0 &  0   \\
  & error $\left[ \% \right]$ & 0.14  & - & -  \\
  \hline
  \hline
  \multirow{3}{*}{$\gammat{nl}{5}$} & numeric  &  -   & 2.1 & 0.22 \\
  & analytic  & -    & 2 & 0.22 \\
  & error  $ \left[ \% \right]$ & -  & 4.9 & 0.7   \\
  \hline
\end{tabular}
\caption{MQS eigenvalues of a sphere and their radiative corrections, obtained numerically by Eqs. \ref{eq:Kappa2b}, \ref{eq:Kappa3}, \ref{eq:kappa5b}, and analytically by Eqs. \ref{eq:MQSte} and \ref{eq:MQStem}. Relative error.}
\label{tab:MQS}
\end{table}

\subsubsection{Point source excitation}

\begin{figure}[!ht]
\centering
\includegraphics[width= \columnwidth]{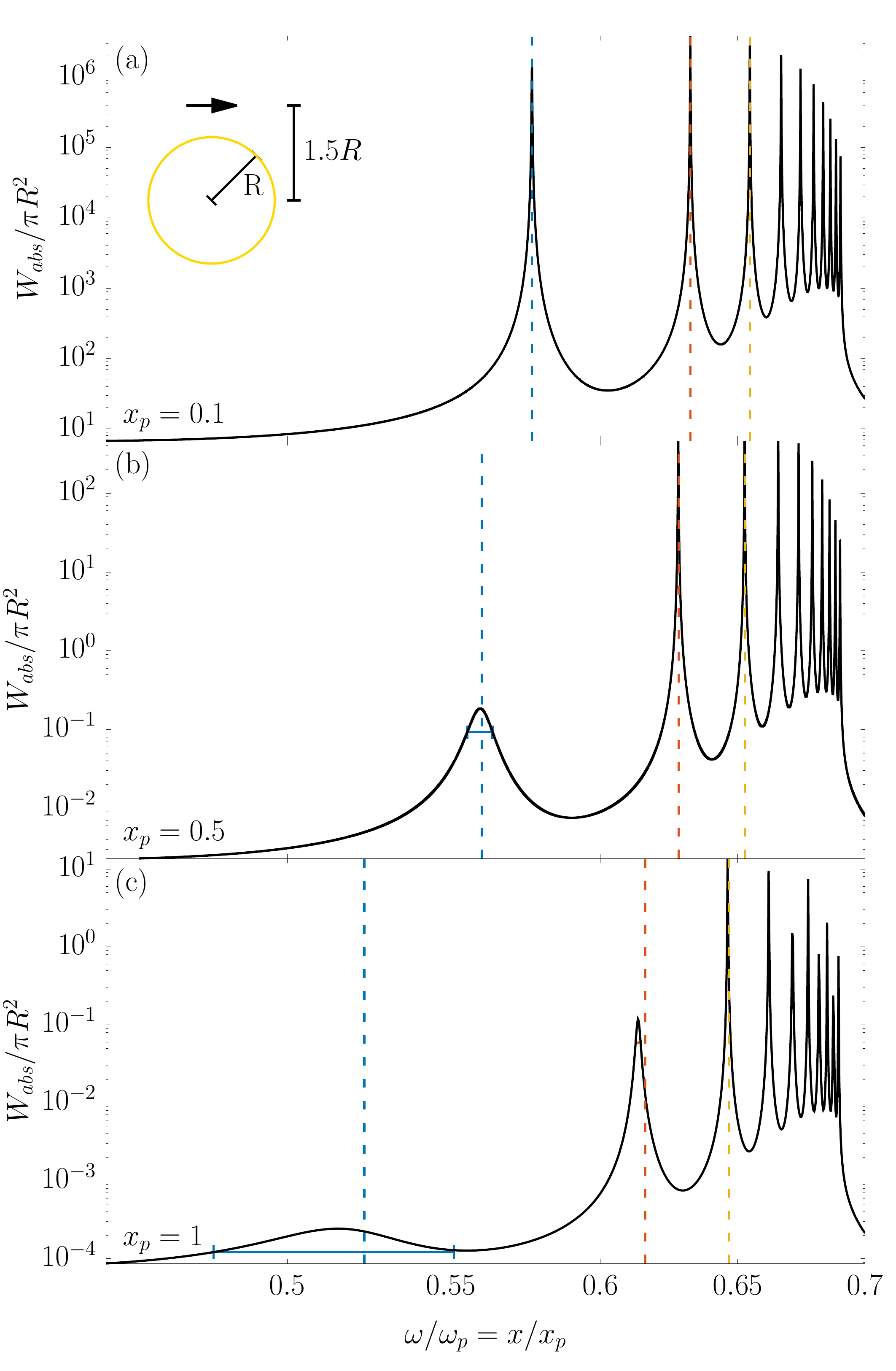}
\caption{Power $W_{abs}$ absorbed by a  Drude metal sphere ($\nu = 10^{-4} \omega_p$)  with radius $R = l_c$ as a function of $\omega / \omega_p = x/x_p$ for $ x_p = 0.1$ (a), $ x_p = 0.5$ (b), $ x_p = 0.5$ (c). The sphere is centered in $\left( 0, 0, 0 \right)$ and it is excited by a point source excitation oriented along $\hat{\mathbf{x}}$, i.e. $\Ei = \mathbf{N}_{e11}^{\left(3\right)}$ at position $\left( 0, 0, 1.5 R \right)$, as in the inset. The resonance frequencies obtained by Eq. \ref{eq:EQSxres2} are shown with vertical dashed lines. Horizontal lines show the FWHM of the broadest peaks.}
\label{fig:Wabs_S_L}
\end{figure}

\begin{figure}[!ht]
\centering
\includegraphics[width=\columnwidth]{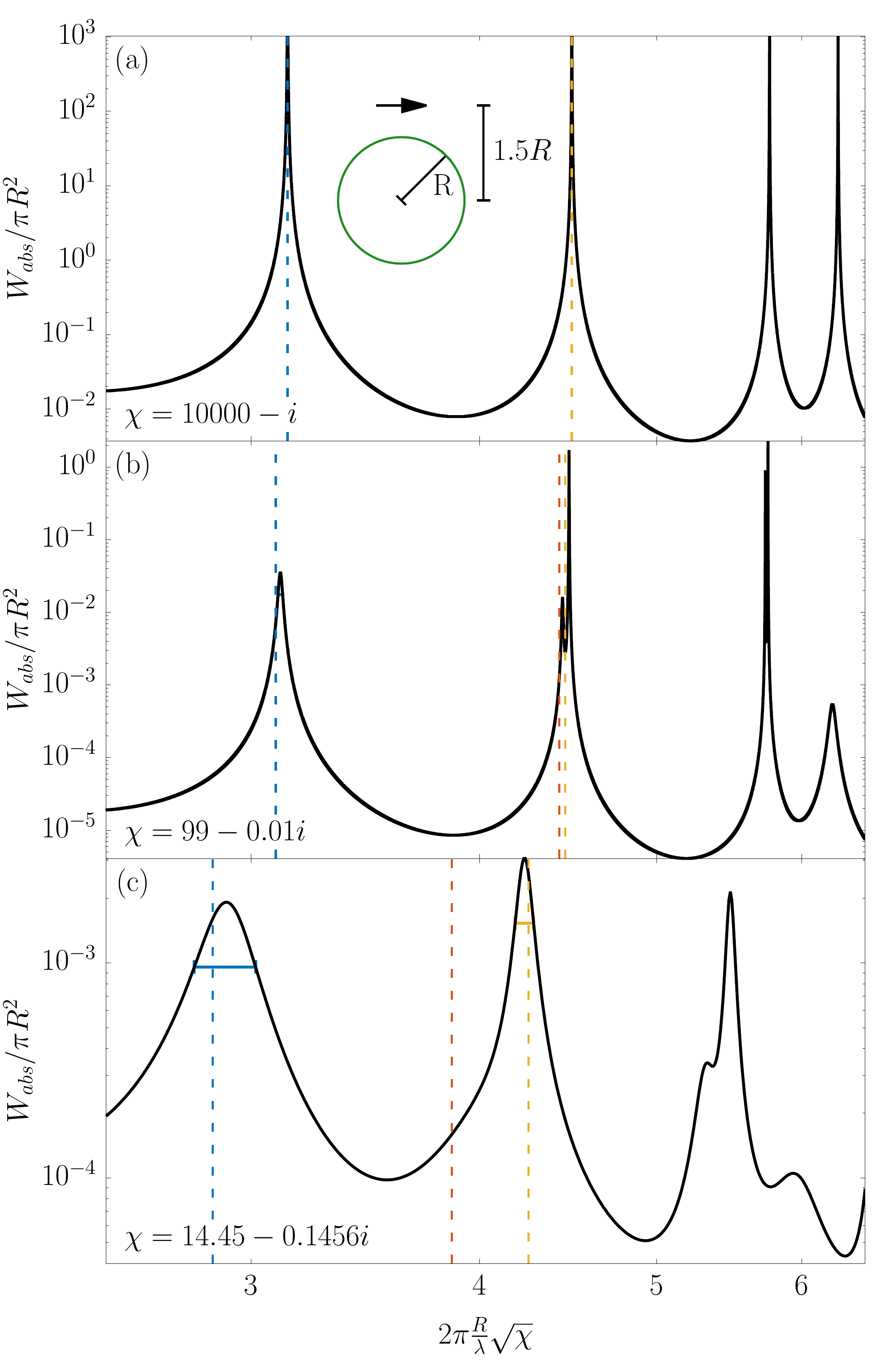}
\caption{Power $W_{abs}$ absorbed by a dielectric sphere with radius $R = l_c$, and susceptibility $\chi =$  (a) $10^4 - 1 i $, (b) $99 - 0.01 i $, (c) $14.45 - 0.1456 i $, as a function of $ x \sqrt{\chi}$. The sphere is centered in $\left( 0, 0, 0 \right)$ and it is excited by point source oriented along $\hat{\mathbf{x}}$, e.g. $\Ei = \mathbf{N}_{e11}^{\left(3\right)}$ at position $\left( 0, 0, 1.5 R \right)$, as in the inset. The first three resonance positions obtained  by Eq. \ref{eq:MQSxres2} are shown with vertical dashed lines. Horizontal lines show the FWHM of the broadest peaks.}
\label{fig:Wabs_S_T}
\end{figure}

The sphere is now excited by a point source. Specifically, the sphere of radius $R$ is centered in the origin, while the point source is oriented along $\hat{\mathbf{x}}$ and it is positioned at $ {\bf r}_d =  \left(0, 0, 1.5R \right)$, namely
\begin{equation}
   \Ei =  \mathbf{N}_{e11}^{\left(3\right)} \left( {\bf r} -  {\bf r}_d  \right),
   \label{eq:PointSource}
\end{equation}
where $ \mathbf{N}_{e11}^{\left(3\right)} $ is a vector spherical wave function of the radiative kind, defined in Eq. \ref{eq:VSWF} of Appendix \ref{sec:VSWF}. Under the same excitation conditions, two different scenarios are investigated. In the first one, shown in Fig.  \ref{fig:Wabs_S_L}, the sphere is  made of a Drude metal, in the second one, shown in Fig.  \ref{fig:Wabs_S_T}, of a high-index dielectric. In both cases low-losses are assumed: this hypothesis is essential for a quantitative comparison between the predicted radiative Q-factor and fractional bandwidth of the peaks, which would have been otherwise dominated by non-radiative losses.
 The  absorbed power spectrum $W_{abs}$  is chosen as physical observable. It is normalized by the geometrical cross section $\pi l_c^2$, and features the expression:
\begin{equation}
 \frac{W_{abs}}{\pi l_c^2} = \frac{x}{2 \pi \zeta_0}   \, \left(- \text{Im} \left\{ \varepsilon_R \right\}  \right) \iiint_{\tOmega} \left\| {\bf E} \left( \tilde{\mathbf{r}}\right) \right\|^2 \dV.
\end{equation}
where $\mathbf{E}$ is the total electric field within the particle, and $\zeta_0 = \sqrt{ \mu_0 / \varepsilon_0}$ is the vacuum characteristic impedance.
In this section, it is calculated by the Mie analytic solution \cite{bohren2008absorption} combined with the translation-addition theorem for vector spherical wave functions (VSWFs)\cite{Wriedt,mishchenko2002scattering}, that is used to translate the VSWF of Eq. \ref{eq:PointSource} into the corresponding VSWFs set centered in the sphere's center. The maximum multipolar expansion order for the Mie solution is assumed to be $20$.

The reason behind the choice of the point source as excitation and of the absorbed power as  physical observable is that without these two hypothesis, some modes may not be excited or probed. 

The radiative shift of the peaks of  $W_{abs}$  and their Q-factors are investigated as the object size increases.  In particular, the plasmonic and dielectric resonance frequencies are compared against the frequencies at which the curve has a peak, denoted as $\hat{\omega}_{h}$. Similarly, the Q-factors of plasmonic and dielectric modes are validated against the corresponding {\it heuristic} Q-factors, given by the ratio of the resonance frequency $\hat{\omega}_{h}$ to the width $ \Delta \omega_{\text{FWHM},h}$ of the resonance curve between two points, at the either side of the resonance, where the ordinate is the half of the maximum absorbed power, namely the full width at half maximum (FWHM) \cite{green1955story}
\begin{equation}
  \hat{Q}_{h} = \frac{\hat{\omega}_{h}}{\Delta \omega_{\text{FWHM},h}}.
  \label{eq:Qh}
\end{equation}
In the $W_{abs}$ spectra of Fig. \ref{fig:Wabs_S_L}, \ref{fig:Wabs_S_T} a segment joining the two ordinates at half maximum is also shown.

{\bf Metal Sphere}. A metal sphere is investigated in Fig.  \ref{fig:Wabs_S_L}, assuming a low-loss Drude metal with $\nu = 10^{-4} \omega_p$. The absorbed power spectra $W_{abs}$ are evaluated as a function of $x/x_p = \omega / \omega_p $ for three different  values of $x_p$: $0.1$ in Fig.  \ref{fig:Wabs_S_L} (a), $0.5$ in Fig.  \ref{fig:Wabs_S_L} (b), $1$ in Fig.  \ref{fig:Wabs_S_L} (c).  It is useful to contextualize the chosen values of $x_p$ to actual materials: for a gold sphere \cite{Maier:03} with $\omega_p \approx 6.79$ T rad/s, they correspond to  $R=4.5nm$ (a), $R=22nm$ (b), and $R=45nm$ (c).  The resonance positions of the first three excited plasmonic modes, which in the quasistatic limit tend to the EQS modes $\Jlo_{e11}$ (electric dipole), $\Jlo_{e12}$ (electric quadrupole), and $\Jlo_{e13}$ (electric octupole), are obtained by Eq. \ref{eq:EQSxres2}, and are shown with vertical dashed lines (blue, red, yellow, respectively).

In the small particle limit, accordingly to Eq. \ref{eq:EQSshiftL2}, the relative frequency shift of any resonance (with respect to its quasistatic position) is a quadratic function of the size parameter at the MQS resonance, whose prefactor depends on the ratio between the second order correction  $\gammal{n}{2}$  and the EQS eigenvalue $\gammalo{n}$. Using the value of corrections shown in Fig. \ref{fig:ModesS_L},  this ratio is found to be significantly larger for  $\Jlo_{e11}$ than for  $\Jlo_{e12}$ and $\Jlo_{e13}$, thus the dipole mode $\Jlo_{e11}$  is expected to exhibit the largest frequency shift.

Moreover, the order $n_i$ of the first non-vanishing imaginary correction is $3$ for the electric dipole mode $\Jlo_{e11}$, $5$ for the electric quadrupole mode $\Jlo_{e12}$, and $7$ for electric octupole mode $\Jlo_{e13}$; thus the dipole mode is also expected to undergo the largest radiative broadening.

In Fig.  \ref{fig:Wabs_S_L} (a), $x_p=0.1$ and the radius $R$ is small compared to the plasma wavelength $\lambda_p$, thus the EQS approximation works well: Eq. \ref{eq:EQSxres0} exactly predicts the occurrence of the $W_{abs}$ peaks.

\begin{table}[h!]
\centering
\begin{tabular}{c|c|c|cc}
\hline
$x_p$ & &  & $\Jlo_{e11}$ & $\Jlo_{e12}$  \\
\hline
\hline
  \multirow{6}{*}{0.5} & theory &  $\omega_n /\omega_p$    &  $0.560$  &   $ 0.628$    \\ \cline{2-5}
                                    &  heuristic &  $\hat{\omega}_n/\omega_p$    &  $0.560$  &   $ 0.632$    \\ \cline{2-5} \cline{2-5}
  &  \multirow{3}{*}{theory}
  &    $\Qlr{n}$ &  68.3 &   1010    \\
  & &$\Qld{n}$ &  5660 &   6280    \\
  & &$\Qlt{n}$ &  67.5 &   3813  \\ \cline{2-5}
  & heuristic & $\Qh{n}$ & 66.6  & 3875  \\  
  \hline
  \hline
  \multirow{6}{*}{1} & theory &  $\omega_n /\omega_p$    &  $0.523$  &   $ 0.616$    \\ \cline{2-5}
                                    &  heuristic &  $\hat{\omega}_n/\omega_p$    &  $0.515$  &   $ 0.6133$    \\ \cline{2-5} \cline{2-5}
                                     &  \multirow{3}{*}{theory} 
 &      $\Qlr{n}$ &  10.5  &  340    \\
 & &  $\Qld{n}$&  5229 &  6159    \\
 & & $\Qlt{n}$  &  10.5 &   320   \\ \cline{2-5}
 & heuristic & $\Qh{n}$ & 7.15 & 340     \\
  \hline
\end{tabular}
\caption{Resonance frequencies ${\omega}_n$ and Q-factors $\Qlr{n}$, $\Qld{n}$, $\Qlt{n}$ of the first plasmonic modes of a Drude metal sphere ($\nu = 10^{-4} \omega_p$), and their heuristic estimates $\hat{\omega}_n$ and $\Qh{n}$.}
\label{tab:Qeqs_S}
\end{table}

In Fig. \ref{fig:Wabs_S_L} (b), $x_p$ is increased to $0.5$, and the $W_{abs}$ peaks undergo a broadening and shift from their quasistatic position, in particular the peaks associated to the electric dipole $\Jlo_{e11}$. Nevertheless, the resonance positions obtained through Eq. \ref{eq:EQSxres2}, which incorporates the radiation corrections, accurately predict the occurrence of the $W_{abs}$ peaks. In Table \ref{tab:Qeqs_S}, the resonance frequencies $\omega_n$ are compared against the corresponding peak positions $\hat{\omega}_n$, while the Q-factors are compared against their heuristic counterparts. In particular, the radiative and non-radiative Q-factors are calculated by Eqs. \ref{eq:eqsQr}, \ref{eq:eqsQnr}, and combined to obtain the total Q-factor by Eq. \ref{eq:eqsQt}. Analytic expressions for the radiative Q-factors of the EQS modes of a sphere are also derived in \ref{eq:Qlsphere}, which agree with the ones provided in Ref. \cite{colas2012mie}. As expected, for $x_p=0.5$ the Q-factor of the dipole mode is the lowest one, and it is limited by radiative losses, unlike all the others. 

Eventually, in Fig. \ref{fig:Wabs_S_L} (c) $x_p = 1$, thus $R$ is comparable to the plasma wavelength $\lambda_p$. The peaks experience a further shift, nevertheless thanks to the radiation corrections, Eq. \ref{eq:EQSxres2} is still able to accurately locate the resonances with an error $< 1.5 \%$, as also shown in Tab. \ref{tab:Qeqs_S}. Also the Q-factors  are predicted with good accuracy. The Q-factor of the dipole and the quadrupole modes are now both dominated by radiative losses.

{\bf High-index sphere.} In Fig. \ref{fig:Wabs_S_T} the power absorbed by a sphere made of  a non-dispersive high-index dielectric with low-losses is investigated as a function of the parameter $x \sqrt{\chi}$. Three  different values of susceptibility $\chi$ are considered , namely $\chi=10^4 - 1 i$, $\chi = 99 - 0.01 i$ $\chi = 14.45  - 0.1456 \, i$.

The expected resonance positions of the three sets of dielectric modes which tend in the quasistatic limit to the MQS modes $\jtTEd{11}$ (magnetic dipole), $\jtTMd{11}$,  and $\jtTEd{21}$ (magnetic quadrupole) are obtained by Eq. \ref{eq:MQSxres2}. They are highlighted  in Fig. \ref{fig:Wabs_S_T} by blue, red, and yellow vertical dashed lines,
respectively.

Accordingly to Eq. \ref{eq:MQSshift}, the resonance frequency shift is a quadratic function of the size parameter at the MQS resonance, whose prefactor depends on the ratio  between the second order correction and the MQS eigenvalue. Using the corrections' values reported in Fig. \ref{fig:ModesS_T}, this ratio is found to be larger for the magnetic dipole $\jtTEd{11}$ than for $\jtTMd{11}$  or $\jtTMd{11}$. Thus this mode is expected to undergo the largest frequency shift.

In Fig.  \ref{fig:Wabs_S_T} (a) it is assumed $\chi=10^4 +i$ and $x \in \left[0.025,0.065  \right] $. Thus the size parameter is very small and the MQS approximation works well:  Eq. \ref{eq:MQSxres0} exactly predicts the occurrence of the $W_{abs}$ peaks.  

In Fig. \ref{fig:Wabs_S_T} (b) it is assumed  $\chi=99 +0.01 i$ and $x \in \left[0.25,0.65  \right]$.  The  peaks position red-shifts against their MQS position, but Eq. \ref{eq:MQSxres2}, which includes the radiation corrections, predicts their occurrence.  In Tab. \ref{tab:QmqsSphere} the expected resonance frequencies and Q-factors are compared against the peak positions and the corresponding heuristic Q-factors.
 In particular, the radiative and non-radiative Q-factors are calculated by Eqs. \ref{eq:mqsQr}, \ref{eq:mqsQnr}, and combined by Eq. \ref{eq:mqsQt}. Analytic expressions of the Q-factors of MQS modes of a sphere are also provided in Eq.  \ref{eq:QmqsTE} and \ref{eq:QmqsTM}.
 
 The agreement of the total Q-factors  with their heuristic counterparts is good. The radiative damping determines the broadening of the magnetic dipole modes, while non-radiative mechanisms play an important role for remaining peaks. 

Eventually, in Fig. \ref{fig:Wabs_S_T} (c),  a silicon sphere with $\chi = 14.45 - 0.1456 i$ is investigated in the range $x \in \left[0.64, 1.6 \right]$. The radius is now comparable to the incident wavelength. The shift of the peaks against the MQS position is significant, but including the radiation corrections, Eq. \ref{eq:MQSxres0} predicts their occurrence with an error $\le 2 \%$.  As shown in Tab. \ref{tab:QmqsSphere} the predicted Q-factors are very close to their heuristic counterparts, and dominated by radiative losses. The second peak is not visible anymore due to the radiative broadening, and its {\it heuristic} Q-factor cannot be evaluated.

\begin{table}[h!]
\centering
\begin{tabular}{c|c|c|c|c|c}
\hline
 $\chi$  & &  & $\jtTEd{11}$  & $\jtTMd{11}$  & $\jtTEd{21}$ \\
\hline
\multirow{6}{*}{$99 -  0.01 \, i$} &  theory & $ x_h \sqrt{\chi} $ &  3.095 & 4.456 & 4.423  \\ \cline{2-6}
&  heuristic & $\hat{x}_{h} \sqrt{\chi}$ &  3.113 &   4.477 & 4.441 \\ \cline{2-6} \cline{2-6}
  & \multirow{3}{*}{theory} & $\Qtr{h}$ &  163 &   579 &   5043 \\
                              &  & $\Qtd{h}$   & $10^4$  & $10^4$ &  $10^4$    \\
                              &  & $\Qtt{h}$   & 163  &  547 & 3352   \\  \cline{2-6}
                              & \multirow{1}{*}{heuristic} & $\Qh{h}$ & 171  & 448 & 3731 \\
\hline
\hline
\multirow{6}{*}{$ 14.45 -  0.145 \, i$} &  theory & $ x_h \sqrt{\chi} $ &  2.860 & 3.863 & 4.255  \\ \cline{2-6}
&  heuristic & $\hat{x}_{h} \sqrt{\chi}$ &  2.907 &   - & 4.237 \\ \cline{2-6} \cline{2-6}
 &  theory & $\Qtt{h} \approx \Qtr{h}$ & 11.60  & 9.6 & 51.7     \\
 & heuristic & $\Qh{h}$  & 13  & - &  43   \\
  \hline
\end{tabular}
\caption{Resonance position $x_h \sqrt{\chi}$ and Q-factors $\Qtr{h}$, $\Qtd{h}$, $\Qtt{h}$ of the first dielectric modes of a dielectric sphere with different values susceptibility $\chi$, and their heuristic estimates $\hat{x}_h \sqrt{\chi}$ and $\Qh{h}$.}
\label{tab:QmqsSphere}
\end{table}

\subsection{Finite Size Cylinder}

A finite-size cylinder with basis radius $R$ and height $H=R$ is now considered. The edges are rounded with a curvature radius of $R/10$. The characteristic length $l_c$ is assumed to be equal to the radius $R$.  The triangular surface mesh used for the calculation of the EQS modes  has $2063$ nodes, and $4122$ triangles. The  hexahedral mesh used for the calculation of MQS modes has $6060$ nodes, $5148$ hexahedra, and $9433$ edges.

\subsubsection{Catalogue of plasmonic resonances}
\begin{figure*}[!ht]
\centering
\includegraphics[width=0.95\textwidth]{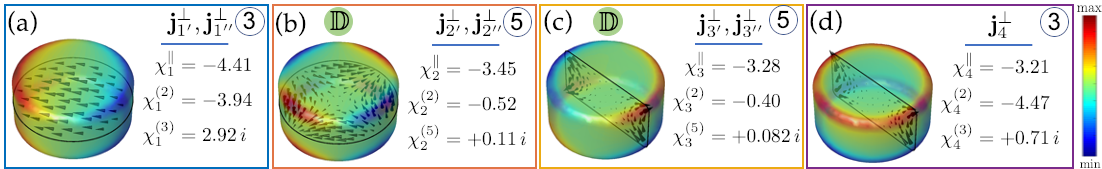}
\caption{Catalogue of plasmonic resonances of a finite-size cylinder with $l_c=R=H$. The electroquasistatic current density modes $\Jlo_{h}$ are ordered according to their eigenvalue $\gammalo{h}$. Their field lines are shown with black arrows on representative planes, and their normal component on $\partial \Omega$ is represented with colors. The second order correction $\gammal{h}{2}$, and by the non-vanishing imaginary correction $\gammal{h}{n}$ of lowest-order $n_i$ are shown on the right of the corresponding box, while the value of $n_i$ is also highlighted on the top-right enclosed in a circle. The {\it dark} modes are also labeled with $\mathbb{D}$.}
\label{fig:ModesC_L}
\end{figure*}

The catalogue of the plasmonic resonances  is shown in Fig. \ref{fig:ModesC_L}. The two degenerate EQS current modes $\left\{\Jlo_{1'}, \Jlo_{1''} \right\}$ have the smallest EQS eigenvalue. They represent two electric dipoles oriented along mutually orthogonal directions, which are  orthogonal to the cylinder's axis, thus they are bright. The second order correction $\gammal{1}{2} = -3.94$ is obtained by Eq. \ref{eq:Chi2b}. The third order correction $\gammal{1}{3} = +2.92 \, i$, given by Eq. \ref{eq:Chi3}, is proportional to the squared magnitude of the electric dipole moment of the EQS modes.

The next EQS modes are two degenerate couples, namely $\left\{ \Jlo_{2'}, \Jlo_{2''}\right\}$, and $\left\{ \Jlo_{3'},\Jlo_{3''}\right\}$ shown in Fig. \ref{fig:ModesC_L} (b)-(c). They are dark, thus the third order corrections $\gammal{2}{3}$ and $\gammal{3}{3}$ vanish. In both cases, the lowest order of the first non-vanishing imaginary correction is the fifth. Accordingly to Eq. \ref{eq:Chi5}, $\gammal{2}{5}$ and $\gammal{3}{5}$ depend on the electric quadrupole tensor of the corresponding EQS mode.

In Fig. \ref{fig:ModesC_L} (d) the EQS mode $\Jlo_{4}$  is shown. This mode is associated to an electric dipole aligned along the cylinder's axis, thus it is bright. The second order correction is $\gammal{4}{2} = -4.47$.   The third order correction is proportional to the squared magnitude of the electric dipole moment of the EQS current mode, its value is $\gammal{4}{3} = +0.71 \, i$.

\begin{figure*}[!ht]
\centering
\includegraphics[width=0.7\textwidth]{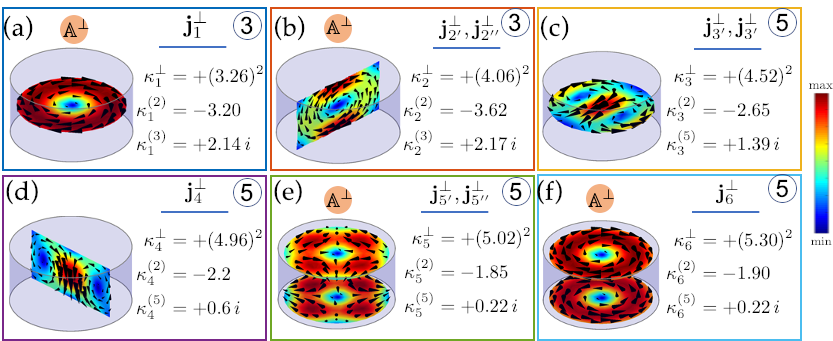}
\caption{Catalogue of dielectric resonances of a finite-size cylinder with $l_c=R=H$.  The magnetoquasistatic current density modes are ordered according to their eigenvalue $\gammato{h}$. The second order correction $\gammat{h}{2}$, and the non-vanishing imaginary correction $\gammat{h}{n}$ of lowest-order $n_i$ are shown on the right of the corresponding box, while the value of $n_i$ is also highlighted on the top-right enclosed in a circle. The current modes generating a transverse vector potential are labeled with the symbol $\mathbb{A}^\perp$.}
\label{fig:ModesC_T}
\end{figure*}

\subsubsection{Catalogue of dielectric resonances}
The catalogue of the dielectric resonances of the finite-size cylinder is presented in Fig. \ref{fig:ModesC_T}.  The mode $\Jto_{1}$ is associated to the lowest MQS eigenvalue, shown in Fig. \ref{fig:ModesC_T} (a), is a magnetic dipole oriented along the cylinder axis. The current mode $\Jto_{1}$ is an  $\mathbb{A}^\perp$-mode, since that the longitudinal component of the vector potential it generates is {\it numerically} negligible. Thus, the magnetostatic interaction energy between the current mode $\Jto_{1}$ and any EQS current mode is zero  and the second order correction $\gammat{1}{2}$ is simply given by Eq. \ref{eq:Kappa2TE}. The third order correction is proportional to the squared magnitude of the magnetic dipole moment of $\Jto_{1}$, accordingly to Eq. \ref{eq:Kappa3}.

The successive couple of degenerate modes $\left\{ \Jto_{2'}, \Jto_{2''} \right\}$, shown in  Fig. \ref{fig:ModesC_T} (b), are magnetic dipoles oriented along two mutually orthogonal  directions, which are both orthogonal to the cylinder's axis.  Analogously to $\Jto_{1}$, they are  $\mathbb{A}^\perp$-modes, and similar considerations apply.

The next degenerate modes, namely $\left\{ \Jto_{3'}, \Jto_{3''} \right\}$, are shown in Fig. \ref{fig:ModesC_T} (c). They are known as  $\text{HEM}_{12\delta}$ within the antenna community \cite{Mongia:94}. Contrarily to the previous modes, these current modes generate a vector potentials with a non-vanishing longitudinal component. Specifically, accordingly to Eq. \ref{eq:Kappa2b}, the second order correction $\gammat{3}{2}$ also depends on the interaction energy between the MQS current modes $ \Jto_{3'}$ and $ \Jto_{3''}$ and the two horizontal EQS current modes $\Jlo_{1'}$, $\Jlo_{1''}$, shown in Fig. \ref{fig:ModesC_L} (a). The coupling with the remaining EQS modes is instead negligible.  Since the modes  $ \Jto_{3'}$ and $ \Jto_{3''}$ have zero magnetic dipole moment, the third order correction vanishes. The imaginary correction of the lowest order is $\gammat{3}{5}$, and it is  determined by the multipolar contribution from the toroidal dipole moment $ \mathbf{P}_{\text{E}2|3}^\perp$ and from the electric dipole moment of the second order correction  $ \mathbf{P}_{\text{E}|3}^{\left( 2 \right)}$.

The next mode is $\Jto_{4}$, also known \cite{Mongia:94} as $\text{TM}_{01\delta}$, and it is shown in Fig. \ref{fig:ModesC_T} (d).  It generates a vector potential with a non-vanishing longitudinal component. In particular, its magnetostatic interaction energy with the EQS mode $\Jlo_{4}$ contributes to the second order correction $\gammat{4}{2}$ of Eq. \ref{eq:Kappa2b}. The mode $\Jto_4$ has zero magnetic dipole moment, thus $\gammat{4}{3}=0$, while its fifth order correction is non-vanishing and determined by the multipoles $ \mathbf{P}_{\text{E}2|4}^\perp$ and $ \mathbf{P}_{\text{E}|4}^{\left( 2 \right)}$.

The subsequent two sets of degenerate modes $ \left\{ \Jto_{5'},  \Jto_{5''} \right\}$ and $ \Jto_{6}$, shown in Fig. \ref{fig:ModesC_T} (e) and (f), are $\mathbb{A}^\perp$-modes. Therefore, the second order correction is given by Eq. \ref{eq:Kappa2TE}. They exhibit zero magnetic dipole moment, thus vanishing third order correction. The fifth order correction $\gammat{5}{5}$ can be calculated by Eq. \ref{eq:kappa5TE}, and originates in both cases from the magnetic quadrupole $\tensor{\mathbf{Q}}_{\text{M}|6}^\perp $ of the MQS mode.

\subsubsection{Point Source Excitation}
\begin{figure}[!ht]
\centering
\includegraphics[width=\columnwidth]{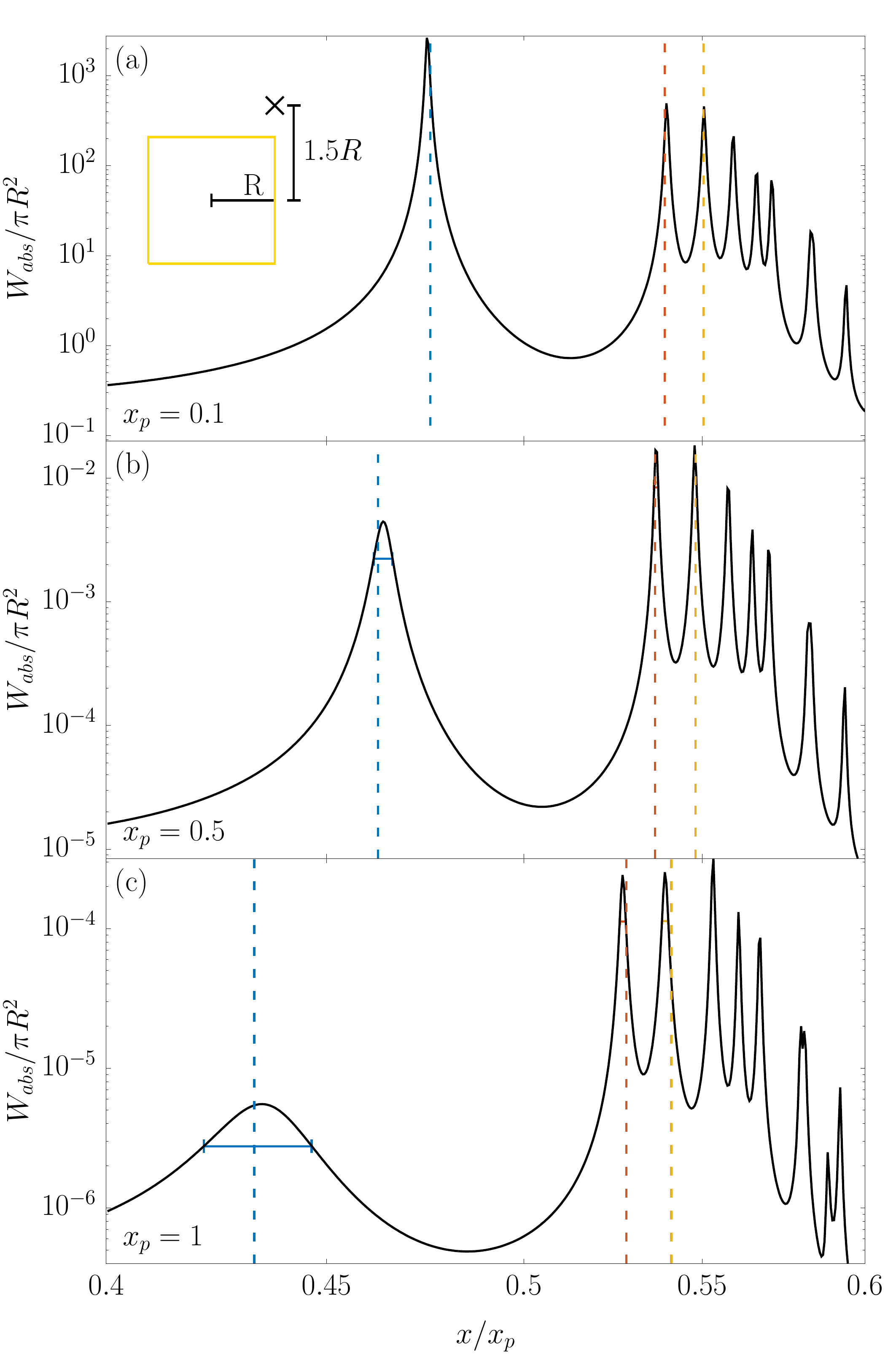}
\caption{Power $W_{abs}$ absorbed by a Drude metal  cylinder ($\nu = 10^{-3} \omega_p$) with radius $R = l_c$ and height $H=R$ as a function of $\omega / \omega_p = x/x_p$ for $ x_p = 0.1$ (a), $ x_p = 0.5$ (b), $ x_p = 0.5$ (c).  The cylinder is centered in $\left( 0, 0, 0 \right)$ and it is excited by a point source oriented along $\hat{\mathbf{y}}$, i.e. $\Ei = \mathbf{N}_{o11}^{\left(3\right)}$ at position $\left( R, 0, 1.5 R \right)$, as in the inset. The resonance frequencies obtained by Eq. \ref{eq:EQSxres2} are shown with vertical dashed lines. Horizontal lines show the FWHM of the broadest peaks.}
\label{fig:WabsC_L}
\end{figure}


Now, the finite-size cylinder is excited by a point source excitation. Specifically, the cylinder of radius $R$ and height $H=R$ is centered in the origin, its axis is oriented along $\hat{\mathbf{z}}$, while the point source is oriented along $\hat{\mathbf{y}}$ and it is positioned at $ {\bf r}_d =  \left( R, 0, 1.5R \right)$, namely
\begin{equation}
   \Ei =  \mathbf{N}_{o11}^{\left(3\right)} \left( {\bf r} -  {\bf r}_d  \right).
\end{equation}
Assuming the same geometry and excitation conditions, two different scenarios are investigated. In the first one, the cylinder is made of a Drude metal, in the second of a high-index dielectric.  The absorbed power spectrum is calculated by an in-house full-wave numerical method based on the surface integral equation method \cite{harrington1993field}, using  a triangular mesh with $1106$ nodes and $2208$ triangles. The frequency shift and the broadening of the resonances are investigated.

{\bf Metal cylinder.} In Fig.  \ref{fig:WabsC_L} the first scenario is investigated, assuming a low loss Drude metal with $\nu = 10^{-3} \omega_p$. The absorbed power $W_{abs}$ is shown as a function of $x/x_p = \omega / \omega_p $ for three different values of $x_p$:   $0.1$ in Fig. \ref{fig:WabsC_L} (a), $0.5$ in Fig. \ref{fig:WabsC_L} (b), and $1$ in Fig. \ref{fig:WabsC_L} (c). For gold cylinders, by assuming, as in Ref. \cite{Maier:03}, $\omega_p \approx 6.79$ Trad/s, the  three values of $x_p$ correspond to  $R=4.5$nm (a), $R=22$nm (b), and $R=45$nm (c). The expected resonance positions of the first three sets of plasmonic modes, which in the quasistatic limit tend to the EQS modes $\left\{ \Jlo_{1'},\Jlo_{1''} \right\}$; $\left\{ \Jlo_{2'}, \Jlo_{2''} \right\}$, and $\left\{ \Jlo_{3'}, \Jlo_{3''} \right\}$, are obtained by Eq. \ref{eq:EQSxres2}, and are shown in Fig.  \ref{fig:WabsC_L} with vertical dashed lines (blue, red, and yellow, respectively).

In Fig.  \ref{fig:WabsC_L} (a), where $x_p = 0.1$,  the radius is significantly smaller than the plasma wavelength and the EQS approximation alone accurately predicts the occurrence of the $W_{abs}$ peaks through Eq. \ref{eq:EQSxres0}. The non-radiative losses cause the broadening of all peaks.
 
In Fig.  \ref{fig:WabsC_L} (b), where $x_p = 0.5$, the absorption peaks begin to shift with respect to the their quasistatic position, in particular the one associated to the horizontal electric dipoles $\Jlo_{1'}$ and $\Jlo_{1''}$. Nevertheless, the equation \ref{eq:EQSxres2}, by including the radiation correction, predicts their occurrence with a relative error less than $0.5\%$, as shown in Table \ref{tab:Qeqs}.
In this table, the Q-factors obtained by Eqs. \ref{eq:eqsQr}, \ref{eq:eqsQnr}, \ref{eq:eqsQt} are also compared against their heuristic counterparts, defined in Eq. \ref{eq:Qh}, and a good agreement is found. In particular, the Q-factor of the the modes $\Jlo_{1'}$ and $\Jlo_{1''}$  is limited by their radiative losses, while  the Q-factors of the modes $\Jlo_{2'}$ $\Jlo_{2''}$ and $\Jlo_{3'}$ $\Jlo_{3''}$ are limited by non-radiative damping mechanisms.  

\begin{table}[h!]
\centering
\begin{tabular}{c|c|c|ccc}
\hline
$x_p$ & &  & $\left\{ \Jlo_{1'}, \Jlo_{1''} \right\}$ & $\left\{ \Jlo_{2'}, \Jlo_{2''} \right\}$ & $\left\{ \Jlo_{3'}, \Jlo_{3''}\right\}$ \\
\hline
  \multirow{6}{*}{0.5} & theory &  $\omega_h /\omega_p$    &  $0.465$  &   $0.5363$ &   $ 0.5480$    \\ \cline{2-6}
                                    & heuristic &  $\hat{\omega}_h /\omega_p$    &  $0.464$  &   $0.5364$ &   $ 0.5478$  \\ \cline{2-6} \cline{2-6}
   &  \multirow{3}{*}{theory}
  &$\Qlr{h}$ &  120 &   2279 &  2596   \\
  & &$\Qld{h}$ &  465 &   536 &  547   \\
  & &$\Qlt{h}$ &   95 &    523 &  537   \\ \cline{2-6}
  & heuristic & $\Qh{h}$ & 100  & 412 &  547  \\  
  \hline
  \hline
    \multirow{6}{*}{1} & theory &  $\omega_h /\omega_p$    &  $0.439 $  &   $0.528$ &   $ 0.541$    \\ \cline{2-6}
                                    & heuristic &  $\hat{\omega}_h /\omega_p$    &  $0.435$  &   $0.527$ &   $ 0.539$  \\ \cline{2-6} \cline{2-6}
                        &  \multirow{3}{*}{theory} 
 &  $\Qlr{h}$ &  17.8  &  768 &   865   \\
 & &  $\Qld{h}$&  439 &  528 & 540   \\
 & & $\Qlt{h}$  &  17.1 &   313 &  332   \\ \cline{2-6}
 & heuristic & $\Qh{h}$ & 17  & 310  & 292   \\
  \hline
\end{tabular}
\caption{Resonance frequencies ${\omega}_h$ and Q-factors $\Qlr{h}$, $\Qld{h}$, $\Qlt{h}$ of the first plasmonic modes of a Drude metal cylinder ($\nu = 10^{-4} \omega_p$) with $R=H$, and their heuristic estimates $\hat{\omega}_h$ and $\Qh{h}$.}
\label{tab:Qeqs}
\end{table}

In Fig. \ref{fig:WabsC_L} (c), where $x_p=1$, the cylinder has radius comparable to the plasma wavelength. Despite the further shift of the peaks, the radiation corrections are still able to correctly locate the resonances, with an error less than $0.5\%$ (see Tab. \ref{tab:Qeqs}). Moreover, Tab. \ref{tab:Qeqs} shows that while the broadening of the first peak arises from the radiative damping, both radiative and non-radiative damping contribute to the broadening of the second and third peak. The total Q-factor is also in good agreement with the heuristic Q-factor in the investigated cases.

\begin{figure}[!ht]
\centering
\includegraphics[width=\columnwidth]{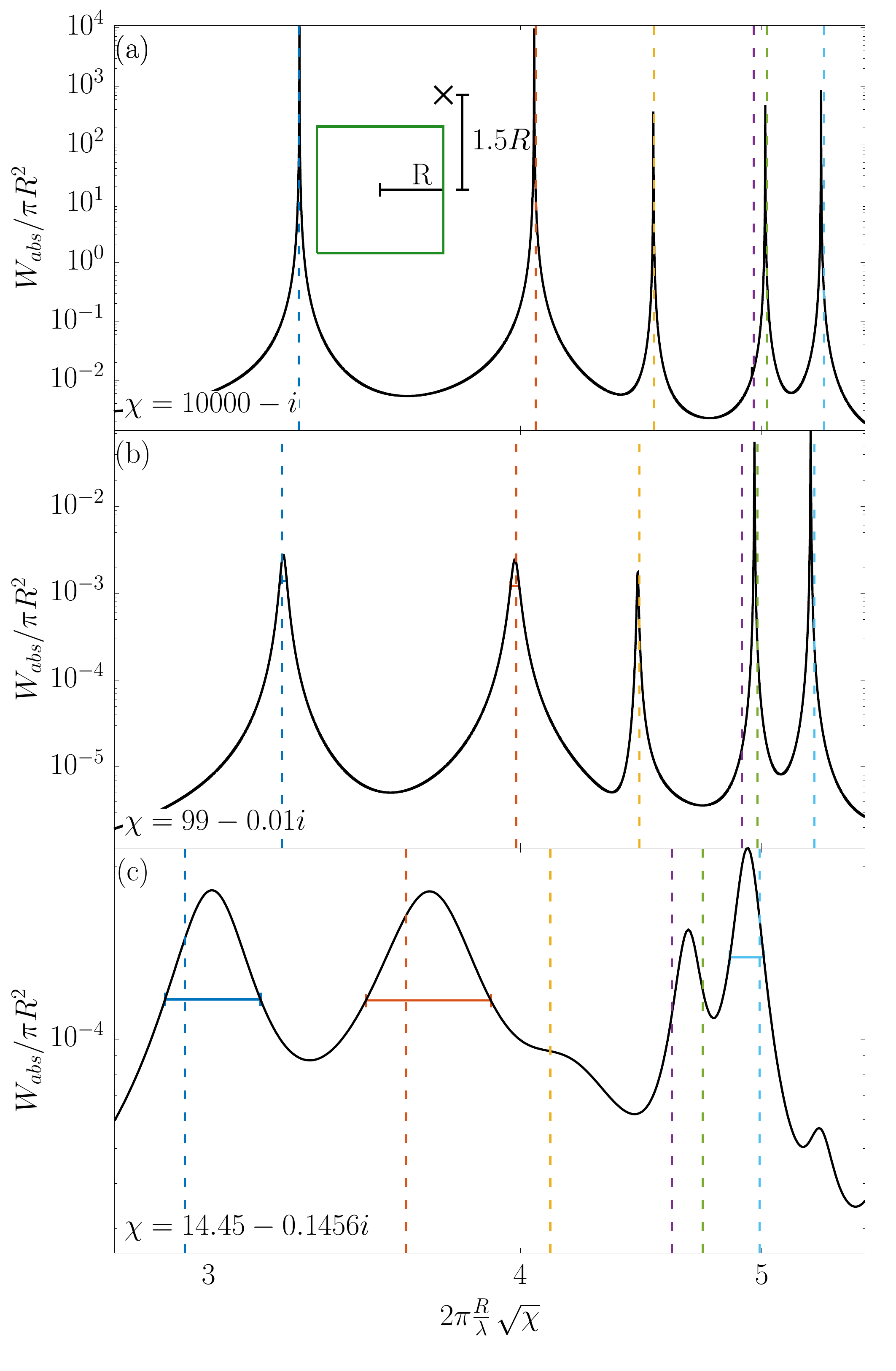}
\caption{Power $W_{abs}$ absorbed by a finite-size cylinder with radius $R = l_c$, height $H=R$, and $\chi=$  (a) $10^4 - 1 i $, (b) $99 - 0.01 i $, (c) $14.45 - 0.1456 i $, as a function of $y = x \sqrt{\chi}$. The cylinder is centered in $\left( 0, 0, 0 \right)$ and it is excited by a point source excitation oriented along $\hat{\mathbf{y}}$, e.g. $\Ei = \mathbf{N}_{o11}^{\left(3\right)}$, at position $\left( R, 0, 1.5 R \right)$.The first three resonance positions obtained by Eq.  \ref{eq:MQSxres2} are shown with vertical dashed lines. Horizontal lines show the FWHM of the broadest peaks.}
\label{fig:WabsC_T}
\end{figure}


{\bf High-index cylinder.} The power absorbed by a cylinder constituted by a non-dispersive high-index dielectric with low-losses is now investigated as a function of $ x \sqrt{\chi}$.
Three different values of $\chi$, namely $\chi =10^4 - 1 i$, $\chi = 99 - 0.01 i$, and $\chi= 14.45 - 0.1456 i$ are considered in Fig. \ref{fig:WabsC_T} (a), (b), and (c), respectively.  The resonance positions obtained by Eq. \ref{eq:MQSxres2}, of the first six sets of dielectric modes, which in the quasistatic limit tend to the MQS modes $\Jto_{1}$, $\left\{ \Jto_{2'}, \Jto_{2''} \right\}$,  $\left\{ \Jto_{3'}, \Jto_{3''}\right\}$, $\left\{ \Jto_{4'}, \Jto_{4''}\right\}$, $\left\{ \Jto_{5'}, \Jto_{5''} \right\}$, and $\Jto_{6}$, are shown with dashed vertical lines (blue, red, yellow, violet, green, and cyan, respectively.)

In Fig.  \ref{fig:WabsC_T} (a) it is assumed that $\chi = 10^4 - 1 i$ and $x \in \left[0.00275,0.055  \right]$. The size parameter is very small and even the MQS approximation alone (without corrections) accurately predicts the occurrence of the peaks by Eq. \ref{eq:MQSxres0}.

Next, in Fig. \ref{fig:WabsC_T} (b), it is assumed that $\chi = 99 - 0.01 i$ and $x \in \left[0.0275,0.55  \right]$. The peaks undergo a shift from their quasistatic positions,  but Eq. \ref{eq:MQSxres2}, by taking into account the second order radiation correction, predicts their occurrence with an error $<0.2\%$. The resonance positions and Q-factors of $\Jto_{1}$, $\left\{ \Jto_{2'}, \Jto_{2''} \right\}$,  and $\Jto_{6}$ are compared in Tab. \ref{tab:Q_C_T} against the  peak positions and heuristic Q-factors. Good agreement is found.

Eventually, in Fig. \ref{fig:WabsC_T} (c), a silicon cylinder with $\chi = 14.45 - 0.1456 i$ is investigated in the range $x \in \left[0.72, 1.44 \right]$. The size parameter is now of the order of one and the peaks undergo a significant shift and broadening. Nevertheless,  the MQS approximation equipped with radiation corrections  is able to predict through Eq. \ref{eq:MQSxres2} the peaks occurrence with an error lower than $2.5\%$, as shown in Tab. \ref{tab:Q_C_T}. In this same table, the corresponding Q-factors, all dominated by radiative losses, are also shown and are very close to their {\it heuristic} counterpart.  The resonance associated to  $\left\{ \Jto_{3'}, \Jto_{3''}\right\}$ (vertical yellow dashed line) only corresponds to a shoulder in the $W_{abs}$ curve due to the radiative broadening, thus it was not possible to define its {\it heuristic} resonance position. Furthermore, the interplay of the modes $\left\{ \Jto_{4'}, \Jto_{4''}\right\}$ and $\left\{ \Jto_{5'}, \Jto_{5''}\right\}$ results in only one $W_{abs}$ peak. For these reasons they are not reported in Tab. \ref{tab:Q_C_T}.

\begin{table}[h!]
\centering
\begin{tabular}{c|c|c|c|c|c}
\hline
$ \chi$  & &  & $\Jto_{1} \Jto_{1'}$ & $\Jto_{2} \Jto_{2'}$  & $\Jto_{6}$ \\
\hline
  \multirow{6}{*}{$99 -  0.01 \, i$} & theory &  $x_h \sqrt{\chi}$    &  $3.209$  &   $3.986$ &   $  5.248$    \\ \cline{2-6}
                                    & heuristic &  $\hat{x}_h \sqrt{\chi}$    &  $3.214$  &   $3.980$ &   $ 5.230 $  \\ \cline{2-6} \cline{2-6}
                              & \multirow{3}{*}{theory} & $\Qtr{h}$ &  147 &   118 &   3117 \\
                              &  & $\Qtd{h}$   & $10^4$  & $10^4$ &  $10^4$    \\
                              &  & $\Qtt{h}$   & 144  &  116 &  2615  \\  \cline{2-6}
                              & \multirow{1}{*}{heuristic} & $\Qh{h}$ & 156  &  126 & 2615 \\
\hline
\hline
  \multirow{4}{*}{$ 14.45 -  0.1456 \, i$}& theory &  $x_h \sqrt{\chi}$    &  $2.935$  &   $3.600$ &   $ 4.989 $    \\ \cline{2-6}
                                    & heuristic &  $\hat{x}_h \sqrt{\chi}$    &  $3.007$  &   $3.677$ &   $ 4.938 $  \\ \cline{2-6} \cline{2-6}
 & theory &  $ \Qtt{h} \approx \Qtr{h}  $  & 10.74  &  8.87  & 32.6    \\
 &  heuristic & $\Qh{h}$ & 11.5  &  8.67  & 31.9  \\
  \hline
\end{tabular}
\caption{Resonance position $x_h \sqrt{\chi}$  and Q-factors $\Qtr{h}$, $\Qtd{h}$, $\Qtt{h}$ of the first dielectric modes of a dielectric cylinder ($H=R$) with different susceptibility $\chi$, and their corresponding heuristic estimates $\hat{x}_h \sqrt{\chi}$and $\Qh{h}$.}
\label{tab:Q_C_T}
\end{table}

\subsection{Triangular Prism}
A triangular prism with basis edge $L$ and height $H=L/2$ is now investigated. The edges and corners are rounded with a curvature radius of $L/20$. The characteristic length is assumed to be equal to half of the edge length, i.e. $l_c = L/2$. The triangular surface mesh used for the calculation of EQS resonances has $2349$ nodes, and $4694$ triangles, while the hexahedral mesh used for the calculation of MQS resonances has $2520$ nodes, $2025$ hexahedra, and $3592$ edges.

\subsubsection{Catalogue of plasmonic resonances}
\begin{figure*}[!ht]
\centering
\includegraphics[width=0.95\textwidth]{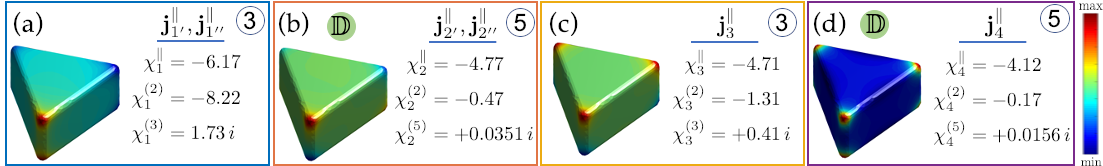}
\caption{ Catalogue of plasmonic resonances of a triangular prism with $l_c=R/2$. The electroquasistatic current density modes $\Jlo_{h}$ are ordered according to their eigenvalue $\gammalo{h}$. Their field lines are shown with black arrows on representative planes, and their normal component on $\partial \Omega$ is represented with colors. The second order correction $\gammal{h}{2}$, and by the non-vanishing imaginary correction $\gammal{h}{n}$ of lowest-order $n_i$. The value of $n_i$ is also highlighted on the top-right of each box.The {\it dark} modes are also labeled with $\mathbb{D}$.}
\label{fig:ModesT_L}
\end{figure*}

The  {\it catalogue} of plasmonic resonances  of a triangular prism is presented in Fig. \ref{fig:ModesT_L}.   The first  two degenerate EQS current modes  are $\left\{ \Jlo_{1'},\Jlo_{1''} \right\}  $, depicted in Fig. \ref{fig:ModesT_L} (a). They  represent two electric dipoles oriented along mutually orthogonal directions, and are indeed bright. The third order correction is proportional to the squared magnitude of their electric dipole moment, accordingly to Eq. \ref{eq:Chi3}.

The next EQS modes  $\left\{ \Jlo_{2'}, \Jlo_{2''} \right\}$ exhibit a quadrupolar character. They are described in Fig. \ref{fig:ModesT_L} (b). They are dark,and for this reason the third order correction $\gammal{2}{3}$ vanishes. Nevertheless, they have a non vanishing electric quadrupole tensor, thus  the fifth order correction $\gammal{2}{5}$  is non-zero accordingly to Eq. \ref{eq:Chi5}.

The successive EQS mode, $\Jlo_3$, is bright and corresponds to a vertical electric dipole. The third order correction $\gammal{3}{3}$ is proportional to the squared magnitude of its electric dipole moment of $\Jlo_3$.

The next EQS mode  $\Jlo_{4}$ exhibits a quadrupolar character. It is shown in Fig. \ref{fig:ModesT_L} (d). It is dark, thus the third order correction $\gammal{h}{3}$ vanishes. Nevertheless, it has a non vanishing electric quadrupole tensor and the fifth order correction $\gammal{4}{5}$  is non-zero and it is given by Eq. \ref{eq:Chi5}.

\begin{figure*}[!ht]
\centering
\includegraphics[width=0.7\textwidth]{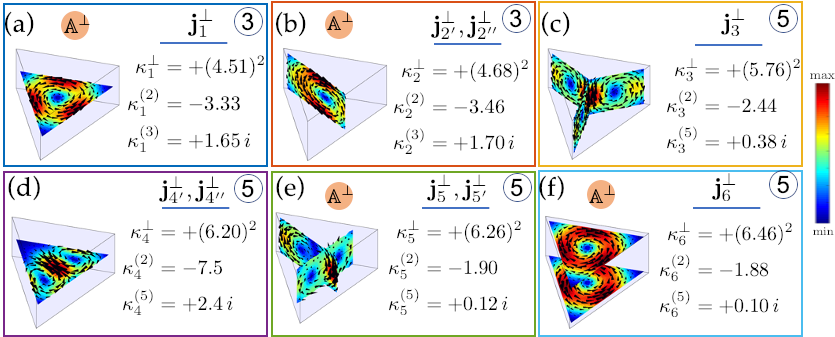}
\caption{Catalogue of dielectric resonances of a triangular prism with $l_c=R/2$.  The magnetoquasistatic current density modes are ordered according to their eigenvalue $\gammato{h}$. The second order correction $\gammat{h}{2}$, and the non-vanishing imaginary correction $\gammat{h}{n}$ of lowest-order $n_i$ are shown on the right of the corresponding box, while the value of $n_i$ is also highlighted on the top-right enclosed in a circle. The current modes generating a transverse vector potential are labeled with the symbol $\mathbb{A}^\perp$.}
\label{fig:ModesT_T}
\end{figure*}

\subsubsection{Catalogue of dielectric resonances}
The {\it catalogue} of dielectric resonances of the finite-size triangular prism is presented in Fig. \ref{fig:ModesT_T}.  The mode $\Jto_{1}$, shown in Fig. \ref{fig:ModesT_T} (a), is associated to the lowest MQS eigenvalue. It is a magnetic dipole oriented along the vertical axis. The current mode $\Jto_{1}$ is an  $\mathbb{A}^\perp$-mode, since the longitudinal part of the vector potential generated by  $\Jto_{1}$ is {\it numerically} negligible. This fact exemplifies that $\mathbb{A}^\perp$-modes can be also found in non-rotationally symmetric objects. Thus the second order correction $\gammat{1}{2}$ is simply given by Eq. \ref{eq:Kappa2TE}.  The third order correction is proportional to the squared magnitude of the magnetic dipole moment of $\Jto_{1}$, accordingly to Eq. \ref{eq:Kappa3}.

The next two degenerate modes $\left\{ \Jto_{2'}, \Jto_{2''} \right\}$ are in-plane magnetic dipoles, one of them is shown in  Fig. \ref{fig:ModesT_T} (b). They are $\mathbb{A}^\perp$-modes with a non-vanishing magnetic dipole moment and the considerations made for $\Jto_{1}$ also apply here.

The next mode $ \Jto_{3}$ is shown in Fig. \ref{fig:ModesT_T} (c).  The vector potential generated by the current mode $\Jto_{3}$ has a non-vanishing longitudinal component. In particular, the magnetostatic interaction energy between $\Jto_{3}$ and the vertical EQS current modes $\Jlo_{3}$, shown in Fig. \ref{fig:ModesT_L} (c), contributes to the second order correction $\gammat{3}{2}$. Since the mode $ \Jto_{3}$ has zero magnetic dipole, the third order correction vanishes. The imaginary correction of the lowest order is $\gammat{3}{5}$, and it is  determined by the effective dipole resulting from the interference between the toroidal dipole moment $ \mathbf{P}_{\text{E}2|3}^\perp$ and the electric dipole moment of the second order correction  $ \mathbf{P}_{\text{E}|3}^{\left( 2 \right)}$. 

The next degenerate modes are $\left\{ \Jto_{4'}, \Jto_{4''} \right\}$, they are described in Fig. \ref{fig:ModesT_T} (d).  They generate a vector potential with a non-vanishing longitudinal component. Specifically, the magnetostatic interaction energy between them and two EQS electric dipole modes $\Jlo_{1'}$,$\Jlo_{1''}$ is non-vanishing and contributes to the second order correction $\gammat{4}{2}$ in Eq. \ref{eq:Kappa2b}. Since $\Jto_{4'}$, $\Jto_{4''}$ have zero magnetic dipole moment, thus $\gammat{4}{3}=0$, while the fifth order correction is non-vanishing and  determined by the multipoles $ \mathbf{P}_{\text{E}2|4}^\perp$ and $ \mathbf{P}_{\text{E}|4}^{\left( 2 \right)}$.

The subsequent two sets of MQS modes, i.e. $\left\{ \Jto_{5'},  \Jto_{5''} \right\}$ and $ \Jto_{6}$, shown in Fig. \ref{fig:ModesT_T} (e) and (f), are $\mathbb{A}^\perp$-modes. Therefore, the second order correction is given by Eq. \ref{eq:Kappa2TE}. These modes exhibit zero magnetic dipole moment, thus vanishing third order correction. The fifth order corrections can be calculated by Eq. \ref{eq:kappa5TE}, and in both cases originates from the magnetic quadrupole of the MQS mode.

\section{Conclusions}

Maxwell's equations provide an exhaustive description of classical electromagnetic phenomena, from the simplest to the most sophisticated ones.  However, in many applications, wave phenomena occurring at short time scales are of no practical concern, and the fields may be described by either their magnetoquasistatic or the electroquasistatic approximation, i.e. the approximations behind the description of capacitors and inductors. This is also the case for light scattering: resonances in metal or high-index objects, assumed much smaller than the vacuum wavelength, may be respectively described by the electroquasistatic or the magnetoquasistatic approximation. Unfortunately, both approximations are unable to predict the frequency-shift and radiative Q-factors, which arise from the coupling with the radiation. 

 In this paper, closed form expressions for the radiation corrections to the real and imaginary parts of both electroquasistatic and magnetoquasistatic eigenvalues and of the corresponding modes are derived. These corrections only depend on the quasistatic current mode distribution.

The expression of the radiation corrections are greatly simplified if the magnetoquasistatic mode generates a transverse vector potential, namely with vanishing normal component to the surface of the object ($\mathbb{A}^\perp$-modes). 

The relative frequency shift of any mode is a quadratic function of the size parameter at the quasistatic resonance, whose prefactor depends on  the ratio between the second order correction and the quasistatic eigenvalue. The {\it radiative} Q-factor is an inverse power function of the size parameter whose exponent is  the order $n_i \ge 3$ of the first non-vanishing imaginary correction, while the prefactor is the ratio between the the static eigenvalue and its $n_i$-th order imaginary correction. Specifically,  the prefactor only depends on the quasistatic eigenvalue and on the multipolar components of the quasistatic modes.

The resonances of a small objects can be then naturally classified in two catalogues of plasmonic and dielectric resonances, containing the essential information to analyze and engineer the electromagnetic scattering from small objects. In these tables the resonances are sorted accordingly to their real quasistatic eigenvalue, and characterized by the second order correction and by the non-vanishing imaginary correction of the lowest order, i.e. $n_i  $.  All the quantities contained in these tables do not depend neither on the size nor on the permittivity, but only on the quasistatic mode morphology.

The introduced expressions for the resonance frequency and Q-factor are successfully validated by predicting the resonance peaks and their broadening in the absorption spectra of a sphere and a finite-size cylinder. Both Drude-metals and high- index dielectric are considered.

\appendix

\section{Multipoles}
\label{sec:Multipoles}
The electric dipole moment $\PE{h}$ of the $h$-th electroquasistatic current mode  $\Jlo_{h}$ is defined as:
\begin{equation}
    \PE{h}	 =  \iiint_{\tOmega} \Jlo_{h} \, \dV =  \oiint_{\partial \tOmega} \left( \Jlo_{h}  \cdot  \n \right) \, \rbt \, \dS,
    \label{eq:Dipole}
\end{equation}
the electric quadrupole tensor  $\QE{h} $ as:
\begin{equation}
   \QE{h} = \iiint_{\tOmega} \mathbf{r}  \, \Jlo_{h} + \Jlo_{h} \, \mathbf{r}   \; \dV =  \oiint_{\tOmega}  \left( \Jlo_{h}  \cdot  \n \right)  \, \rbt \rbt \, \dS,
   \label{eq:Quadrupole}
\end{equation}
with respect to standard definitions of electric multipoles \cite{Jackson}, the prefactor $1 / \left( j \omega \right) $ is omitted here.

 The magnetic dipole moment $\PM{h}$ of the $h$-th magnetoquasistatic current mode $ \Jto_{h} $ is defined as
\begin{equation}
    \PM{h} = \frac{1}{2} \iiint_{\tOmega} \rbt \times \Jto_{h} \,  \dV,
    \label{eq:DipoleM}
\end{equation}
the toroidal dipole $\mathbf{P}_{\text{E}2|h}$ as
\begin{equation}
     \mathbf{P}_{\text{E}2|h}^\perp=\frac{1}{6} \iiint_{\tOmega} \rbt \times \Jto_{h} \times \rbt \, \dV,
     \label{eq:ToroidalM}
\end{equation}
and the magnetic quadrupole $\tensor{\mathbf{Q}}_{\text{M}|h}$ tensor as
\begin{equation}
\tensor{\mathbf{Q}}_{\text{M}|h}^\perp =\frac{1}{3} \iiint_{\tOmega} [(\rbt \times \Jto_{h}) \rbt+\rbt (\rbt \times \Jto_{h})] \, \dV.
	\label{eq:QuadrupoleM}
\end{equation}

\section{Quasistatic modes of a sphere}
 \label{sec:AnalyticSphere}
A sphere of radius $R$ is considered, and a characteristic dimension is assumed equal to the radius $l_c=R$. The formulas of radiation corrections presented in this section can be  extrapolated by perturbing the denominators of the Mie coefficients in the neighborhood of their EQS and MQS resonances, and they can also be directly obtained from the Padè expansion of the Mie coefficients found by D. C. Tzarouchis and A. Sihvola in Refs. \cite{Tzarouchis:16,tzarouchis2018light}.
 \begin{figure*}[!ht]
\centering
\includegraphics[width=0.8\textwidth]{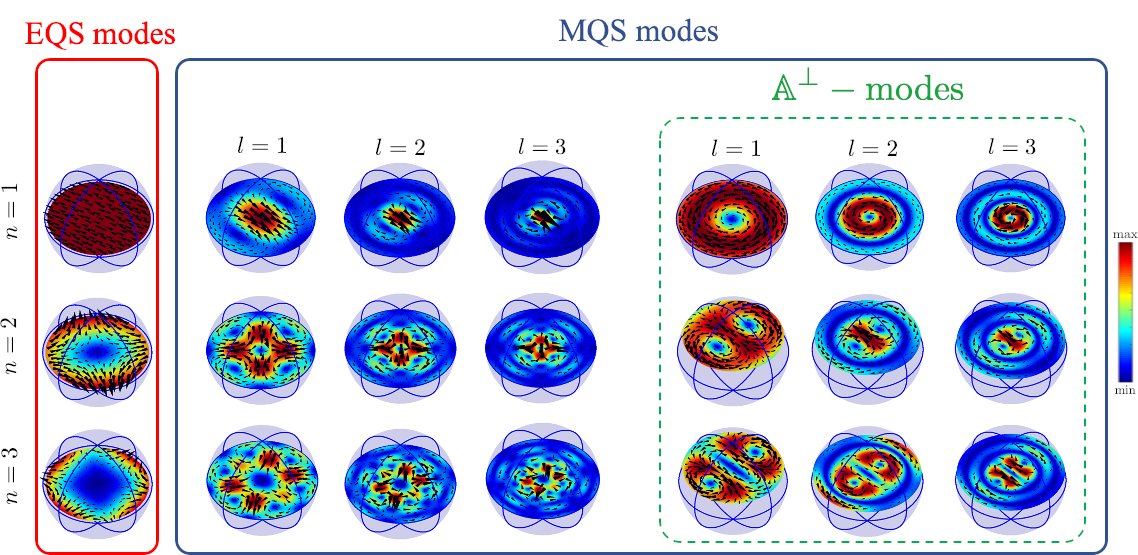}
\caption{Quasistatic modes of a sphere. Electroquasistatic modes $\mathbf{j}_{e1n}^\parallel$ with $n=1,2,3$. The magnetoquasistatic modes are divided into two subsets:  the TM MQS modes $\mathbf{j}_{e1nl}^{\perp \text{TM}} \left( \tilde{r}, \theta, \phi \right)$ with $n=1,2,3$ and $l=1,2,3$, the TE modes ($\mathbb{A}^\perp$-modes) $\mathbf{j}_{o1nl}^{\perp \text{TE}} \left( \tilde{r}, \theta, \phi \right)$ with $n=1,2,3$ and $l=1,2,3$.  }
\label{fig:Modes_Sphere}
\end{figure*}
\subsection{Electroquasistatic modes}
In particular, the EQS eigenvalues  of a sphere and their radiative corrections are \cite{Tzarouchis:16,tzarouchis2018light}:
\begin{subequations}
\begin{align}
   \label{eq:EQSsphere0}
   \gammalo{n}&= - \frac{2n+1}{n}, \\
   \label{eq:EQSsphere2}
   \gammal{n}{2} &= - \frac{2}{n^2} \frac{ \left( n+1 \right)\left( 2 n+1 \right) }{\left( 3 + 2 n \right)\left( 2 n - 1 \right)}, \\   
     \label{eq:EQSsphere3} 
    \gammal{n}{2n+1} &= + i \, \frac{ \left( n+1 \right) }{ \left[ n \left( 2 n - 1 \right)!! \right]^2}, \
\end{align}
\label{eq:EQSsphere}
\end{subequations}
where $n=1,2,3,\ldots$, and $\left(2n  - 1\right)!! = 1 \times 3 \times 5 \times \cdots \times \left(2n  - 1\right)$. {Each eigenvalue  $\gammalo{n}$ is associated to a set of $2n+1$ degenerate current modes with $m=0,1,2,\ldots,n$ and different parity, whose analytic expression are: 
\begin{widetext}
\begin{equation}
\begin{aligned}
\mathbf{j}_{\substack{ e \\ o}mn}^\parallel &=  \frac{1}{\sqrt{\alpha_{mn}}} \left[    \left( \begin{array}{c} \cos  m \phi  \\ \sin  m \phi  \end{array} \right) n P_n^m \left( \cos{\theta} \right)  \, \boldsymbol{\hat{ r}} + 
  \frac{d P_n^m \left( \cos{\theta} \right)}{d \theta} \cos{m \varphi} \,  \boldsymbol{\hat{\theta}}  +   \left( \begin{array}{c} - \sin  m \phi  \\ + \cos  m \phi  \end{array} \right) m  \frac{P_n^m \left( \cos{\theta} \right) }{\sin{\theta}}  \, \boldsymbol{\hat{\phi}} \right] , \\
 \alpha_{mn} &= 2 \pi \left( \delta_m +  1 \right)   \frac{ (m + n)! (n + 1)^{n - 1} }{ \left(2 n + 1 \right)  \left(n - m \right)! n^{n - 2}},
 \end{aligned}
 \label{eq:EQSmode}
\end{equation}
\end{widetext}
where the subscripts $e$ and $o$ denote even and odd, $P_n^m$ are the associated Legendre functions of the first kind of degree $n$ and order $m$ as defined and normalized in Ref. \cite{bohren2008absorption}. The prefactor $\alpha_{mn}$ guarantees that $\left\| \mathbf{j}_{\substack{e \\ o} mn}^\parallel  \right\| = 1$. As an example, the EQS modes $\mathbf{j}_{e1n}^\parallel$ with $n=1,2,3$ are shown in Fig. \ref{fig:Modes_Sphere}.

The radiative quality factor of plasmonic resonances is obtained by using Eq. \ref{eq:EQSsphere2} into Eq. \ref{eq:eqsQr}:
\begin{equation}
    \Qlr{n} = n  \frac{ \left[ \left( 2n + 1 \right)!! \right]^2}{ \left( n + 1 \right) \left( 2 n  + 1 \right)}  \left( \frac{1}{x_h} \right)^{2n+1}.
    \label{eq:Qlsphere}
\end{equation}
For instance, for electric dipole, quadrupole, octupole is
\begin{equation}
   \Qlr{1} = \frac{3}{2} \frac{1}{x_1^3}, \quad  \Qlr{2} =  \frac{30}{x_2^5}, \quad \Qlr{3} =  \frac{1181}{x_3^7}.
\end{equation}
Eq. \ref{eq:Qlsphere} coincide with the formulas provided by G. Colas des Francs in Ref. \cite{colas2012mie}.

\subsection{Magnetoquasistatic modes}
The MQS modes are divided in two sets. The first set is composed by the current modes  which have no radial component. Since the corresponding electric field has the same property, these modes are called {\it transverse electric} or TE modes. They also generate a vector potential, which has non-vanishing normal component to the particle-surface, so they are also $\mathbb{A}^\perp$-modes. The second set of current modes generate a magnetic field  with  vanishing radial component so they are called TM modes. 


\subsubsection{TE MQS Modes}
The first set of MQS modes is made by the set of TE MQS modes which coincides with the $\mathbb{A}^\perp$-mode of a sphere. Their eigenvalues and the corresponding corrections are \cite{Tzarouchis:16,tzarouchis2018light}:
\begin{subequations}
\begin{align}
   \label{eq:MQSte0}
    \kappa_{nl}^{\text{TE} \, \perp} &=  \left( z_{n-1,l} \right)^2, \\
   \label{eq:MQSte2}
   \kappa_{nl}^{\text{TE} \, \left( 2 \right) }&= -\frac{2n+1}{2n-1}, \\    
   \label{eq:MQSte3}
  \kappa_{nl}^{\text{TE} \, \left( 2 n + 1\right) }&= + i \frac{2}{\left[\left( 2n-1\right)!! \right]^2}, 
\end{align}
\label{eq:MQSte}
\end{subequations}
where $z_{nl}$ denotes the $l$-th zero of the spherical Bessel function $j_n$.   Each eigenvalue  $ \kappa_{nl}^{\text{TE} \, \perp}$ is associated to a set of $2n+1$ degenerate current modes with $m=0,1,2,\ldots,n$  and with even and odd parity, whose analytic expression is 
\begin{widetext}
\begin{equation}
	\begin{aligned}
    \mathbf{j}_{\substack{e\\o}mnl}^{\perp \text{TE}} \left( \tilde{r}, \theta, \phi \right) &= \frac{1}{\sqrt{\beta_{mnl}}} \left[ m \left( 
   \begin{array}{cc} - \sin  m \phi  \\  + \cos  m \phi  \end{array} \right)  \frac{P_n^m \left( \cos \theta  \right)}{\sin \theta} \,  \boldsymbol{\hat{\theta}}  - \left( \begin{array}{cc} \cos  m \phi  \\ \sin  m \phi  \end{array} \right) \frac{d P_n^m \left( \cos \theta \right) }{d\theta} \boldsymbol{\hat{\varphi}} \right] j_n \left(  z_{n-1,l}  \, \tilde{r} \right) \\
\beta_{mnl} &= 2 \pi \left( \delta_m + 1 \right)  \frac{ n \left( n + 1 \right)  \left( m + n \right)! }{ 2 \left( 2 n + 1 \right) \left( n - m \right)! } \,  j_n^2 \left(  z_{n-1,l}  \right)
\end{aligned}
 \label{eq:MQSmodeTE}
\end{equation}
\end{widetext}
where the prefactor $\beta_{mnl}$ guarantees that $\left\|  \mathbf{j}_{\substack{e\\o}mnl}^{\perp \text{TE}} \right\| = 1$. As an example, the odd MQS modes $\mathbf{j}_{o1nl}^{\perp \text{TE}} \left( \tilde{r}, \theta, \phi \right)$ with $m=1$, $n=1,2,3$, and $l=1,2,3$ are shown in Fig. \ref{fig:Modes_Sphere}.

The radiative quality factor of the TE MQS modes is obtained by combining Eq. \ref{eq:MQSte} with \ref{eq:mqsQr}:
\begin{equation}
    \Qtr{nl} =  \frac{ \left[   z_{n-1,l}  \left( 2n-1\right)!! \right]^2}{2 }   \left( \frac{1}{x_{nl}} \right)^{2n+1}.
    \label{eq:QmqsTE}
\end{equation}

For instance, for the magnetic dipole and quadrupole:
\begin{equation}
  \Qtr{1,1} = \frac{\pi^2}{2} \frac{1}{x_1^3} \approx \frac{4.9}{x_{1,1}^3}; \; \Qtr{2,1} = \frac{9 \, z_{1,1}^2}{2} \frac{1}{x_{2,1}^5} \approx \frac{90.7}{x_{2,1}^5}.
\end{equation}

\subsubsection{TM MQS modes}
The second set of MQS modes is made by the set of TM MQS modes. Their eigenvalues and the corresponding corrections are \cite{Tzarouchis:16,tzarouchis2018light}:
\begin{subequations}
\begin{align}
   \label{eq:MQStem0}
   \kappa_{nl}^{\text{TM} \, \perp} &=  \left( z_{nl} \right)^2, \\
    \label{eq:MQStem2}
   \kappa_{nl}^{\text{TM} \, \left( 2 \right)}  &= - \frac{n+2}{n}, \\    
    \label{eq:MQStem3}
    \kappa_{nl}^{\text{TM} \, \left( 2n + 3 \right) }&= + i \frac{2}{n^2  \left[\left( 2n-1\right)!! \right]^2},
\end{align}
\label{eq:MQStem}
\end{subequations}
Each eigenvalue  $\kappa_{nl}^{\text{TM} \, \perp} $ is associated to a set of $2n+1$ degenerate current modes with $m=0,1,2,\ldots,n$  and with even and odd parity whose analytic expression is 
\begin{widetext}
\begin{multline}   
\mathbf{j}_{\substack{e\\o} mnl}^{\text{TM} \, \perp} \left( \tilde{r}, \theta, \phi \right) = \frac{1}{\sqrt{\gamma_{mnl}}}
    \left\{     \left( \begin{array}{cc} \cos  m \phi  \\ \sin  m \phi  \end{array} \right) n \left( n + 1 \right) P_n^m \left( \cos \theta \right) 	\frac{ j_n \left( z_{nl} \,\tilde{r} \right)}{z_{nl} \,\tilde{r}} \boldsymbol{\hat{r}} \right. \\     + \left. \left( \begin{array}{cc} \cos  m \phi  \\  \sin  m \phi  \end{array} \right) \frac{d P_n^m \left( \cos \theta \right) }{d\theta}  \frac{1}{ z_{nl} \,\tilde{r} } \frac{d}{dr} \left[ \tilde{r} j_n \left(  z_{nl}  \tilde{r} \right) \right]   \, \boldsymbol{\hat{\theta}}  +
    m \left( \begin{array}{cc} - \sin  m \phi  \\ + \cos  m \phi  \end{array} \right) \frac{ P_n^m \left( \cos \theta \right) }{\sin \theta}    \frac{1}{ z_{nl} \,\tilde{r} } \frac{d}{dr} \left[ \tilde{r} j_n \left(  z_{nl}  \tilde{r} \right) \right] \,\boldsymbol{\hat{\varphi}} \right\}
 \label{eq:MQSmodeTM}
\end{multline}

\begin{equation}
\gamma_{mnl} =  \pi  \left(1 +  \delta_m \right)  \frac{n \left(n + 1\right) \left(n + m\right)! }{ \left(2 n + 1\right)^2 \left(n - m\right)!}  \left[ \left(1 + n\right) j_{n-1}^2 \left(  z_{nl} \right) +  
    n j_{n+1}^2 \left(  z_{nl} \right) \right]
\end{equation}
\end{widetext}
where the prefactor $\gamma_{mnl} $ guarantees that $\left\| \mathbf{j}_{\substack{e\\o} mnl}^{\text{TM} \, \perp}  \right\| = 1$.
where the subscripts $e$ and $o$ denote even and odd, and $P_n^m \left( \cdot \right)$ are the associated Legendre function of the first kind of degree $n$ and order $m$. 

The MQS modes $\mathbf{j}_{e1nl}^{\perp \text{TM}} \left( \tilde{r}, \theta, \phi \right)$ with $n=1,2,3$ and $l=1,2,3$ are shown in Fig. \ref{fig:Modes_Sphere}.

The radiative quality factor of the TE MQS modes is obtained by combining Eq. \ref{eq:MQStem} with \ref{eq:mqsQr}:
\begin{equation}
    Q_{nl}^{r \, \text{TM} \, \perp}  =  \frac{  \left[  n \, z_{n,l} \left( 2n-1\right)!! \right]^2}{2 }   \left( \frac{1}{x_{nl}} \right)^{2n+3}
    \label{eq:QmqsTM}
\end{equation}
As an example the toroidal  electric dipole mode has radiative Q factor:
\begin{equation}
  Q_{1,1}^{r \, \text{TM} \, \perp} = \frac{z_{1,1}^2}{2} \frac{1}{x_{1,1}^5} \approx  \frac{10.1}{x_{1,1}^5}.
\end{equation}

\section{Point Source Excitation}
\label{sec:VSWF}
Following  \cite{bohren2008absorption}, the explicit expression of the vector spherical wave function  $\mathbf{N}_{\substack{ e \\ o}11}^{\left( 3 \right)}$ of the radiative kind is

\begin{widetext}
\begin{multline}
\mathbf{N}_{\substack{ e \\ o}11}^{\left( 3 \right)} =  - \left( \begin{array}{cc} \cos  \phi  \\ \sin  \phi  \end{array} \right) \left( -\frac{i}{\left( k_0 r \right)^3} + \frac{1}{\left( k_0 r \right)^2} \right) 2 \sin{\theta} e^{- i k_0 r } \, \hat{\bf e}_r  \\ 
+   \left( \begin{array}{cc} \cos  \phi  \\ \sin  \phi  \end{array} \right)  \cos{\theta}  \left( - \frac{i}{\left( k_0 r \right)^3}+ \frac{1}{\left( k_0 r \right)^2} + \frac{i}{\left( k_0 r \right)} \right) e^{- i k_0 r }  \, \hat{\bf e}_ \theta \\
 +  \left( \begin{array}{cc} - \sin   \phi  \\ +\cos \phi  \end{array} \right) \left(  - \frac{i}{\left( k_0 r \right)^3} + \frac{1}{\left( k_0 r \right)^2} +  \frac{i}{\left( k_0 r \right)}\right) e^{- i k_0 r } \,  \hat{\bf e }_\phi
 \label{eq:VSWF}
\end{multline}
\end{widetext}
where the subscripts $e$ and $o$ denote even and odd. The Eq. \ref{eq:VSWF} is proportional to the electric field radiated  in vacuum by the electric dipole $\mathbf{p} = \hat{\mathbf{x}}$ when the parity index is $e$ and by $\mathbf{p} = \hat{\mathbf{y}}$ when  the parity index is odd.  Those field can be exactly obtained by multiplying Eq. \ref{eq:VSWF}  by the  factor $ i \frac{1}{4\pi\varepsilon_0} \frac{ \omega^3}{c^3}$.

\end{document}